\RequirePackage{fix-cm}
\documentclass[smallextended]{svjour3}       
\smartqed  
\usepackage{graphicx}
\usepackage{textcomp}
\usepackage[super]{nth}
\usepackage{subfigure}
\usepackage{booktabs}
\usepackage{multirow}

\begin{document}

\title{On Usage of Non-Volatile Memory as Primary Storage for Database Management Systems
\thanks{The research leading to these results has received funding from the European Union\textquotesingle s 7th Framework Programme under grant agreement number 318633, the Ministry of Science and Technology of Spain under contract TIN2015-65316-P, and a HiPEAC collaboration grant awarded to Naveed Ul Mustafa.
}
}

\titlerunning{Naveed Ul MUSTAFA et al. On Usage of NVM as Primary Storage for DBMS}        

\author{Naveed Ul Mustafa \and Adri\`{a} Armejach  \and
        Ozcan Ozturk \and  Adri\'{a}n Cristal 
        \and \\ Osman S. Unsal
}


\institute{N. Ul Mustafa \at
              Department of Computer Engineering, Bilkent University, Ankara, Turkey. \\
              Present Address: Department of Computer Engineering, TED University, Ankara, Turkey.\\
              \email{naveed.ul.mustafa0083@gmail.com}           
           \and
           A. Armejach \at
              Barcelona Supercomputing Center (BSC), Barcelona, Spain and Universitat Polit\`{e}cnica de Catalunya (UPC), Barcelona, Spain.
           \and O. Ozturk \at
              Department of Computer Engineering, Bilkent University, Ankara 06800, Turkey.
           \and A. Cristal \at 
              Barcelona Supercomputing Center (BSC), Barcelona, Spain.
           \and O. S. Unsal \at
              Barcelona Supercomputing Center (BSC), Barcelona, Spain.
}

\date{Received: date / Accepted: date}

\maketitle

\begin{abstract}
This paper explores the implications of employing non-volatile memory (NVM) as primary storage for a data base management system (DBMS). We investigate the modifications necessary to be applied on top of a traditional relational DBMS to take advantage of NVM features. As a case study, we modify the storage engine (SE) of PostgreSQL enabling efficient use of NVM hardware. We detail the necessary changes and challenges such modifications entail and evaluate them using a comprehensive emulation platform. Results indicate that our modified SE reduces query execution time by up to 45\% and 13\% when compared to disk and NVM storage, with average reductions of 19\% and 4\%, respectively. Detailed analysis of these results shows that while our modified SE is able to access data more efficiently, data is not close to the processing units when needed for processing, incurring long latency misses that hinder the performance. To solve this, we develop a general purpose library that employs helper threads to prefetch data from NVM hardware via a simple API. Our library further improves query execution time for our modified SE when compared to disk and NVM storage by up to 54\% and 17\%, with average reductions of 23\% and 8\%, respectively.
\keywords{Non volatile memory \and Relational DBMS \and Storage engine}
\end{abstract}

\section{Introduction}
\noindent The traditional design of a DBMS assumes a memory hierarchy where datasets are stored in disks. Disks are a cheap and non-volatile storage medium suitable for storing large datasets. However, they are extremely slow for data retrieval. To hide their high data-access latency, DRAM is used as an intermediate storage between disks and the processing units.

DRAM is orders of magnitude faster than a disk, and with increasing DRAM chip densities and decreasing memory prices, relational in-memory DBMSs have become increasingly popular \cite{abraham2013scuba,farber2012sap,lindstrom2013ibm,barber2011blink,PeletonLink,pavlo2017self}. Significant components of in-memory DBMSs, like index structures \cite{larson2011sql,zhang2016reducing}, recovery mechanisms for system failure \cite{ongaro2011fast,diaconu2013hekaton}, and commit processing \cite{lee2001single} are tailored towards the usage of main memory as primary storage. An in-memory DBMS assumes that all data fits in main memory. However, for some applications, the database size can grow larger than DRAM capacity \cite{debrabant2014prolegomenon}. At the same time, inherent physical limitations related to leakage current and voltage scaling limit the further scaling of DRAM \cite{mandelman2002challenges,driskill2010latest}. The capacity problem can be resolved by distributing the database across multiple machines but at the cost of performance degradation \cite{debrabant2014prolegomenon}. Furthermore, due to the volatile nature of DRAM, in-memory DBMSs still use a large pool of disks to provide a form of persistent storage for critical or non redundant data \cite{abraham2013scuba,melnik2010dremel,plattner2011sanssoucidb,sikka2012efficient}.

Non-volatile memory (NVM) is an emerging storage class technology that features persistency as well as significantly faster access latencies than hard disks, with read latencies on the same order of magnitude as DRAM \cite{arulraj2017build}. It also offers byte-addressability like DRAM and higher density \cite{qureshi2009scalable,andrei2017sap}. Prominent NVM technologies are PC-RAM \footnote{PC-RAM: Phase Change Random Access Memory} \cite{raoux2008phase}, STT-RAM \footnote{STT-RAM: Spin Transfer Torque Random Access Memory} \cite{driskill2010latest}, and R-RAM \footnote{R-RAM: Resistive Random Access Memory} \cite{strukov2008missing}. With read latency close to that of DRAM, especially in case of PC-RAM and R-RAM \cite{arulraj2015let,chang2012limits}, NVM technologies are a good candidate to improve the performance of decision support systems (DSS), which are dominated by read-only queries on vast datasets \cite{hagmann2002real}. DSS are computer technology solutions that can be used to facilitate complex decision making \cite{shim2002past}. They convert business information into tangible results \cite{chaudhuri2001database} to help executives take knowledge-based decisions.

A DBMS design should take into account the characteristics of NVM to benefit from its features. Simple ports of a traditional DBMS - designed to use disks as the primary storage medium - to 
NVM will show improvement due to the lower access latencies of NVM. However, adapting a DBMS to fit NVM characteristics can offer a number of benefits beyond lower access latencies. This adaptation requires modifications in the storage engine as well as other components of a DBMS \cite{arulraj2017build} in order to take advantage of NVM features.

In this paper, we study the implications of employing NVM in the design of a DBMS. Our main contributions can be summarized as follows:
\begin{itemize}
 \item We discuss and provide insights on the different available options when including NVM into the memory hierarchy of current systems.
 
 \item We focus on investigating the necessary changes and challenges when modifying an existing, well-tested, widely used, and robust traditional DBMS to benefit from NVM features. As a case study, we selected PostgreSQL which is ranked as the \nth{4} most popular DBMS \cite{DBRanking}.
 
 \item 
 Our modifications aim at providing fast access to data by bypassing the slow disk interfaces while maintaining all the functionalities of a robust DBMS such as PostgreSQL. Our modified SEs target read-dominant DSS queries, providing performance improvements by minimizing data movement operations in a traditional disk-based DBMS.
 
 \item We evaluate our proposed modified SEs of PostgreSQL using a comprehensive emulation platform and the TPC-H~\cite{council2008tpc} benchmark. In addition, we also evaluate an unmodified version of PostgreSQL using both a high-end solid state disk and the emulated NVM hardware.
 
 \item We identify and quantify performance bottlenecks that appear when employing NVM hardware. To further improve the performance of our NVM-enabled SEs, we design and implement a general purpose data prefetching library based on helper threads that tackles the identified performance bottlenecks.
\end{itemize}

Experimental results show that our modified SEs are able to reduce the kernel execution time, where file I/O operations take place, from around 10\% to 3\% on average. In terms of wall-clock query execution time, our modifications improve performance by around 19\% and 4\% on average when compared to unmodified PostgreSQL on disk and on NVM storage, respectively. We find that the performance of our modified SE is limited by the fact that data is not close to the processing units when needed for query processing since it is directly accessed from NVM hardware. This leads to long latency user-level cache misses that decrease the improvements achieved by avoiding expensive data movement operations.

We employ a known technique like helper-threads for data prefetching in the context of a NVM-based database design to resolve the data readiness problem we identify. We implement a general purpose data prefetching library that exposes a simple API allowing the creation of a user-specified number of helper threads that seamlessly prefetch data for a given address and memory size with minimum interference to the computation thread. We investigate different thread mapping schemes with and without hyper-threading. Experimental results show that when using our modified SE with the prefetching library, the kernel execution time drops to 0.05\% on average as compared to around 10\% for unmodified PostgreSQL. In terms of wall-clock query execution time, performance improves by up to 17\% when compared to unmodified PostgreSQL with NVM storage, with 8\% average improvement.

The remainder of this paper is organized as follows:
Section~\ref{sec:background} provides background on features of NVM and
the system software needed to enable its usage. Section~\ref{sec:Implications}
elaborates on the different design choices to integrate NVM into the memory hierarchy
of a computing platform running a DBMS. It also lists
required modifications for a traditional DBMS to leverage NVM features as primary storage. Section~\ref{sec:CaseStudy} explains the
read-write architecture of PostgreSQL and our proposed modifications to
implement NVM-aware SEs for PostgreSQL as a case study. Section~\ref{sec:methodology} describes our methodology to emulate NVM hardware. We evaluate our proposed SEs against baselines in Section~\ref{sec:evaluation}. Section~\ref{sec:library} describes the data prefetching library along with the helper thread mapping schemes we employ in this work, while Section~\ref{sec:library-evalualtion} evaluates the library. Section~\ref{sec:RelatedWork} describes related work on NVM, DBMS and data prefetching. We conclude this paper in Section~\ref{sec:conclusion}.

\section{Background}
\label{sec:background}
\noindent In this section, we first describe in detail the properties of NVM technologies, highlighting the implications these might have in the design of a DBMS. We then describe currently available NVM hardware and  system software to manage NVM.

\subsection{Characteristics of NVM}
\noindent \textbf{Data access latency:} Read latencies for NVM technologies will certainly be significantly lower than those of conventional disks. However, since NVM devices are still under development, sources quote varying read latencies. For example, the read latency for STT-RAM ranges from 1 to 20ns, and PC-RAM is expected to be around 50ns ~\cite{arulraj2015let,wang2013low,perez2010non}. Nonetheless, read latency of some NVM technologies is expected to be similar to that of DRAM ~\cite{mittal2016survey,arulraj2015let,wang2013low,chang2012limits,arulraj2016write,oukid2014sofort,chatzistergiou2015rewind}, which is typically around 60ns.

PC-RAM and R-RAM are reported to have a higher write latency compared to DRAM, but STT-RAM also outperforms DRAM in this regard ~\cite{arulraj2015let,wang2013low}. However, the write latency is typically not in the critical path, since it can be tolerated by using buffers ~\cite{qureshi2009scalable}.

\noindent\textbf{Density:} NVM technologies provide higher densities than DRAM, which makes them a good candidate to be used as main memory as well as primary storage, particularly in embedded systems~\cite{huang2012register}. For example, PC-RAM provides 2 to 4 times higher density as compared to DRAM~\cite{qureshi2009scalable}. Future NVMs are expected to have higher capacity and better scalability than DRAM \cite{oukid2015instant,chakrabarti2014atlas,zhang2015study,viglas2014write}
, and it is expected to scale to lower technology nodes as opposed to DRAM.

\noindent\textbf{Endurance:} The maximum number of writes a memory cell can withstand is lower for most NVM 
technologies when compared to DRAM ~\cite{qureshi2009scalable,zhou2009durable}. Specifically, PC-RAM, R-RAM, 
and STT-RAM have projected endurances of $10^{10}$, $10^{8}$, and $10^{15}$ respectively;  as compared to 
$10^{16}$ for DRAM ~\cite{arulraj2015let}. On the other hand, NVMs exhibit higher endurance than flash 
memory technologies ~\cite{wang2013low}.

\noindent\textbf{Energy consumption:} Since NVM does not need a refresh cycle to maintain data states in memory cells like a  DRAM, 
they are more energy efficient. A main memory designed using PC-RAM technology consumes significantly lower per access write energy as compared to DRAM~\cite{zhou2009durable}. Other NVM technologies also have similar lower energy consumption per bit when compared to DRAM~\cite{arulraj2015let,perez2010non}.

In addition to the features listed above, NVM technologies also provide byte-addressability like DRAM and persistency like disks. Due to these features, NVMs are starting to appear in embedded and energy-critical devices and are expected to play a major role in future computing systems. Companies like Intel and Micron have launched the 3D XPoint memory technology, which features non-volatility \cite{3DXPoint}. Intel has also introduced new instructions to support the usage of persistent memory at the instruction set architecture (ISA) level~\cite{intel2016architecture}.

\subsection{Available NVM hardware}\label{NVDIMM}

While NVM hardware has been available in recent years, it has mainly been used to implement Solid State Disks (SSD) using the NVM Express (NVMe) interface. This technology is not suitable as a DRAM replacement due to its endurance and latency properties. Researchers have been anticipating the arrival of Dual Inline Memory Modules (DIMM) based on NVM to substitute traditional DRAM DIMMs for a long time. Recently, in April 2019, Intel has released its 3D Xpoint DIMM based on NVM technology~\cite{hirofuchi2019preliminary}.

\begin{figure}
\centering
\includegraphics[width=.5\textwidth]{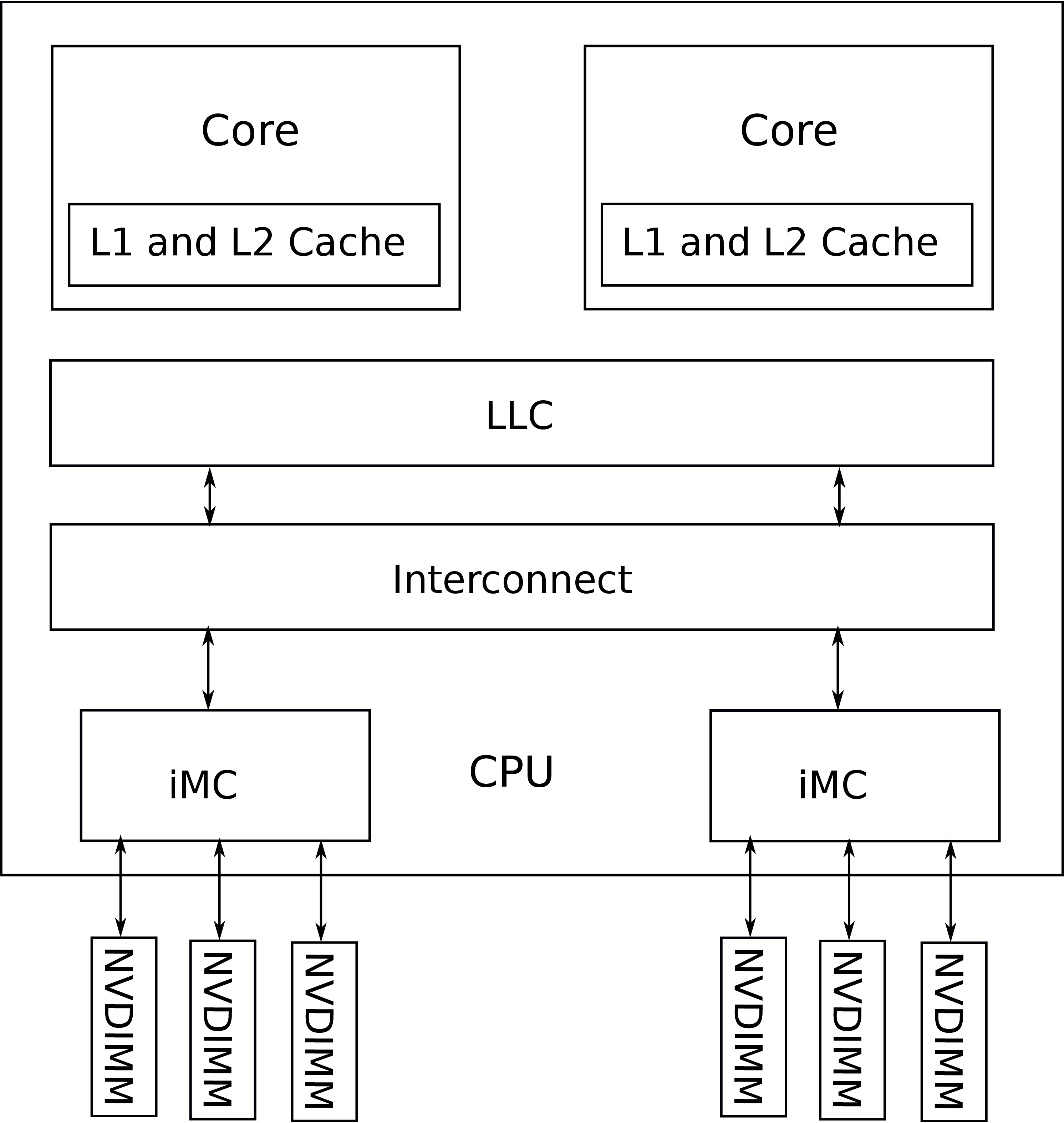}
\caption{Overview of 3D XPoint Memory\textquotesingle s interface with CPU}
\label{3DXPointInterface}
\end{figure}

As shown in Fig.~\ref{3DXPointInterface}, 3D Xpoint DIMMs connect to the memory bus and communicate with a processor through the integrated memory controller (iMC). Each iMC can connect to up to three DIMMs~\cite{peng2019system,yang2019empirical}. A non-standard protocol, DDR-T, is followed for communication between processors and 3D XPoint DIMM. Intel\textquotesingle s Cascade Lake processors are yet the only ones to support 3D Xpoint memory.  As each processor supports two iMCs, six DIMMs are supported in total~\cite{peng2019system,yang2019empirical,izraelevitz2019basic}.

3D XPoint DIMMs can operate in two different modes: Memory and App Direct. Memory mode uses NVM to expand main memory without providing the feature of persistency, while regular DRAM serves as a cache for NVM. In this mode, operating system and CPU see the NVM memory as a volatile extension of main memory. In App Direct mode, NVM is used as a separate persistent memory and does not use DRAM as cache. App Direct mode provides an application direct access to data residing in NVM without interference of the operating system and with byte-addressability. However, it requires an NVM-aware file system to allocate, name, and access persistent data.

\subsection{System software for NVM}

Using NVM as primary storage necessitates modifications not only in application software but also in system software in order to 
take advantage of NVM features. A traditional file system (FS) accesses the storage through a block layer. If a disk is replaced by NVM without any modifications in
the FS, the NVM storage will still be accessed at block level granularity. Hence, we will not be able to take advantage of the byte-addressability feature of NVM. 

For this reason, researchers have developed purpose-built file systems such as NOn Volatile memory Accelerated (NOVA) \cite{xu2016nova} and persistent memory file System (PMFS) ~\cite{dulloor2014system,githubPMFS}. Both file systems expose NVM to an application by providing direct access through a memory map (mmap) interface. As this work uses PMFS, we discuss it in more detail below.

PMFS is an open-source POSIX compliant FS developed by Intel Research. It offers two key features in order to facilitate usage of NVM.  
First, PMFS does not maintain a separate address space for NVM. In other words, both main memory and NVM use the same address space. This implies that there is no need to copy data from NVM to DRAM to make it accessible to an application. A process can directly access file system protected data stored in NVM at byte level granularity.

\begin{figure}
\centering
\includegraphics[width=100mm]{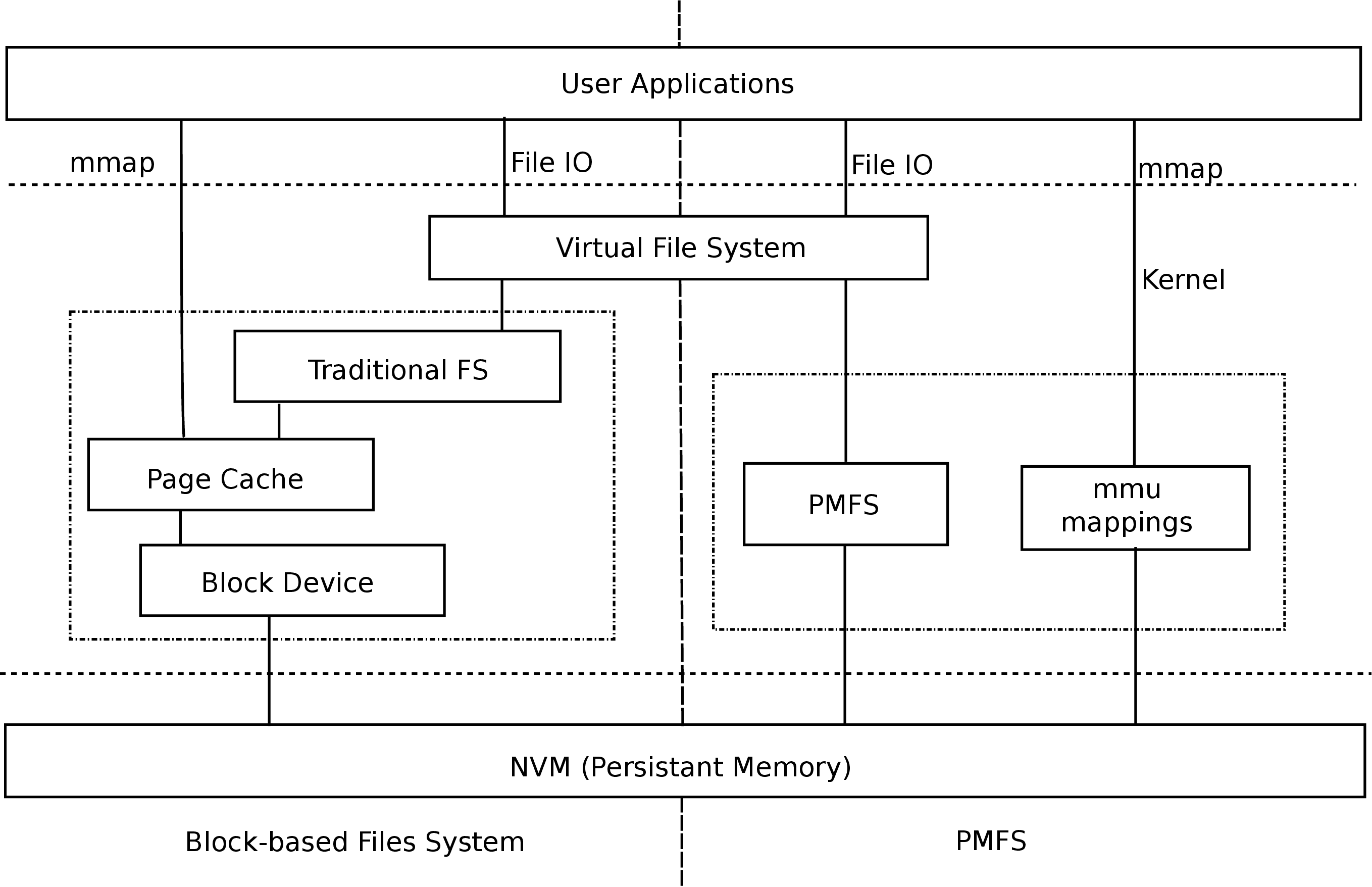}
\caption{Comparison of traditional FS and PMFS. ``mmap'' refers to the system call for memory mapped I/O operation. 
``mmu'' is the memory management unit responsible for address mappings}
\label{Fig1}
\end{figure}

Second, in a traditional FS stored blocks can be accessed in two ways: (i) file I/O and (ii) memory mapped I/O. PMFS implements file I/O in a similar way to a traditional FS. However, the implementation of memory mapped I/O differs. In a traditional FS, memory mapped I/O would first copy pages to DRAM~\cite{dulloor2014system} from where an application can examine those pages. PMFS avoids this copy overhead by mapping NVM pages directly into the address space of a process. Fig.~\ref{Fig1} from~\cite{dulloor2014system} compares a traditional FS with PMFS.


\section{Design Choices}
\label{sec:Implications}
\noindent In this section, we discuss the possible memory hierarchy designs when including NVM in a system. We also discuss the high-level modifications necessary in a traditional disk-optimized DBMS in order to take full advantage of NVM hardware.
\subsection{Memory Hierarchy Designs for an NVM-Based DBMS}

\begin{figure} 
\centering     
\subfigure[Traditional design]{\label{Fig2a}\includegraphics[width=27mm]{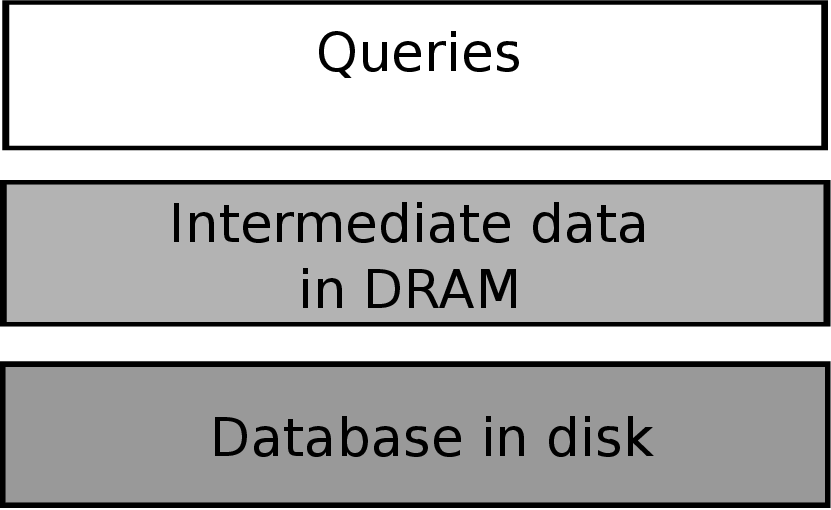}}
\subfigure[All-in-NVM]{\label{Fig2b}\includegraphics[width=27mm]{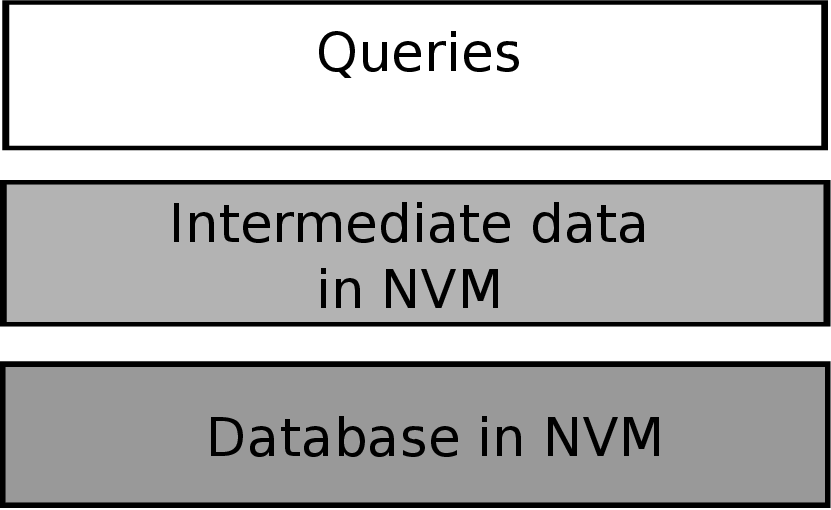}}
\subfigure[NVM-Disk]{\label{Fig2c}\includegraphics[width=27mm]{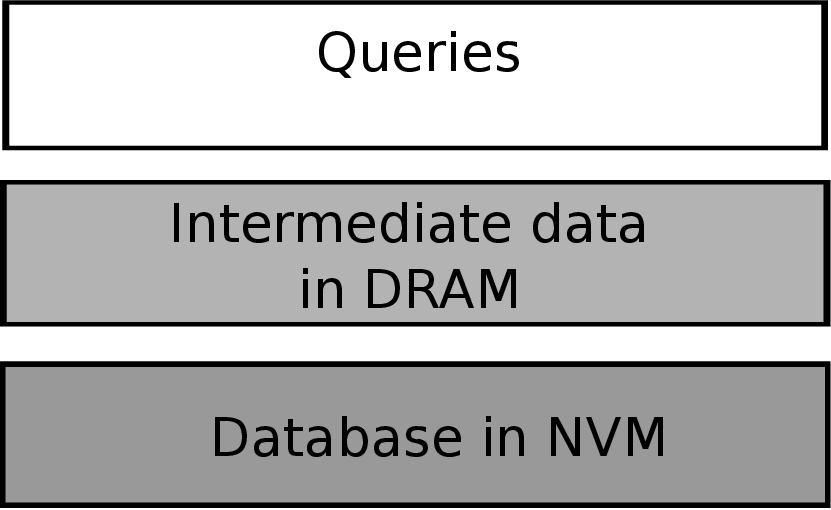}}
\caption{NVM placement in the memory hierarchy of a computing system}
\label{Fig2}
\end{figure}

With features of byte-addressability, low latency and high capacity, NVM has the potential to replace traditional disks as well as main memory \cite{chang2012limits}. Fig.~\ref{Fig2} shows different options that might be considered when including NVM into the system. Fig.~\ref{Fig2a} depicts a traditional approach, where the intermediate state - including logs, data buffers, and partial query state - is stored in DRAM to hide disk latencies for data that is currently in use; while the bulk of the relational data is stored in a  disk.

Given the favorable characteristics of NVM over the other technologies, an option might be to replace both DRAM and disk storage 
using NVM (Fig.~\ref{Fig2b}). However, such a drastic change would require a complete redesign of current operating systems and 
application software. In addition, NVM technology is still not mature enough in terms of endurance to be used as a DRAM replacement. 
Hence, we advocate for a platform that still has a layer of DRAM memory, like \cite{kimura2015foedus}, where the disk is completely or partially replaced using NVM, 
as shown in Fig.~\ref{Fig2c} (NVM-Disk). 

Using this approach, we can retain the programmability of current systems by still having a layer of DRAM, thereby exploiting DRAM's fast read and write access latencies for temporary data structures and application code. In addition, it allows the possibility to directly access the bulk of the database relational data by using a file system such as PMFS, taking full advantage of NVM technology, which allows the system to leverage NVM's byte-addressability and to avoid API overheads~\cite{huang2014nvram} present in current FSs. Unlike an in-memory DBMS, such a setup does not need large pools of DRAM since temporary data is orders of magnitude smaller than the actual relational database stored in NVM. We believe this is a realistic scenario for future systems integrating NVM, with room for small variations such as NVM alongside DRAM to store persistent temporary data structures, or having traditional disks to store cold data.

As explained in Section~\ref{NVDIMM}, 3D XPoint memory can operate in two different modes. Memory mode is similar to a traditional design (Fig.~\ref{Fig2a}), as the 3D XPoint DIMMS are not considered as persistent memory but as the actual DRAM address space. In this mode the DRAM DIMMs are transparently used as a cache for the 3D XPoint DIMMs. This does not change the system view from an application's point of view, and makes sense if one wants to use 3D XPoint DIMMs as if they were large-capacity DRAM DIMMs. However, the App Direct mode would fit into the NVM-Disk (Fig.~\ref{Fig2c}) category. Applications still see DRAM DIMMs as a layer of volatile memory, but can also directly access the 3D XPoint DIMMs via mmap interfaces that enable byte-addressability. The system we consider and later evaluate would be based on a setup similar to that offered by 3D XPoint\textquotesingle s  App Direct mode.


\subsection{Potential Modifications in a Traditional DBMS}~\label{modList}
Using a traditional disk-based database with NVM storage will not take full advantage of NVM's features. Some important components of the DBMS need to be modified or removed when using NVM as a primary storage. 

\noindent\textbf{Avoid the block level access:} Traditional design of a DBMS uses a disk as a primary storage. Since disks favor sequential accesses, database systems hide disk latencies by issuing fewer but larger disk accesses in the form of a data block~\cite{schindler2002track}. 

Unfortunately, block level I/O causes extra data movement. For example, if a transaction updates a single byte of a tuple, it still needs to write the whole block of data to the disk. On the other hand, block level access provides good data locality.
 
Since NVM is byte-addressable, we can read and write only the required byte(s). However, reducing the data retrieval granularity down to a byte level eliminates the advantage of data locality altogether. A good compromise is to reduce the block size in such a way that the overhead of the block I/O is reduced to an acceptable level, while at the same time the application benefits from some degree of data locality. 
 
\noindent\textbf{Remove internal buffer cache of DBMS:} DBMSs usually maintain an internal buffer cache. Whenever a tuple is to be accessed, first its disk address has to be calculated. If the corresponding block of data is not found in the internal buffer cache, then it is read from disk and stored in the internal buffer cache \cite{debrabant2013anti}. 
 
This approach is unnecessary in an NVM-based database design. If the NVM address space is made visible to a process, then there is no need to copy data blocks. It is more efficient to refer to the tuple directly by its address. However, we need an NVM-aware FS, such as PMFS, to enable direct access to the NVM address space by a process.

\subsection{Discussion}
NVM provides the promising features of persistency, like disk storage; and byte-addressability, like DRAM. However, NVMs
have certain limitations such as lower endurance compared to DRAM \cite{arulraj2015let} and a disparity between the read and write latencies \cite{pelley2014memory}. 
Furthermore, different NVM technologies differ from each other in term of these features \cite{arulraj2015let}.

A storage engine aiming to improve decision support
system (DSS) queries can be designed by taking advantage
of the common features of persistency and byte-addressability.
Since DSS queries are read dominant and perform a relatively
negligible number of write operations, the design should
not be influenced or sensitive to different endurance and write
latencies found across NVM technologies. Furthermore, NVM technologies
are projected to provide read latencies similar to DRAM \cite{mittal2016survey,arulraj2015let,wang2013low,chang2012limits}.
Therefore, reading data directly from NVM storage should be comparable in terms
of access latency to reading application data stored in DRAM.

Usage of NVM as primary storage can also impact
other components of a DBMS besides those mentioned
in Section \ref{modList}. For example, if internal buffers are not
employed and all updates are materialized directly into the
NVM address space then the need and criticality of the redo
log can be relaxed \cite{huang2014nvram}. However, the undo log will still be
needed to recover from a system failure. These important aspects are
out of the scope of this work and we will focus on storage engine modifications.

\begin{figure*}  
\centering     
\subfigure[PostgreSQL storage engine]{\label{Fig3a}\includegraphics[width=38mm]{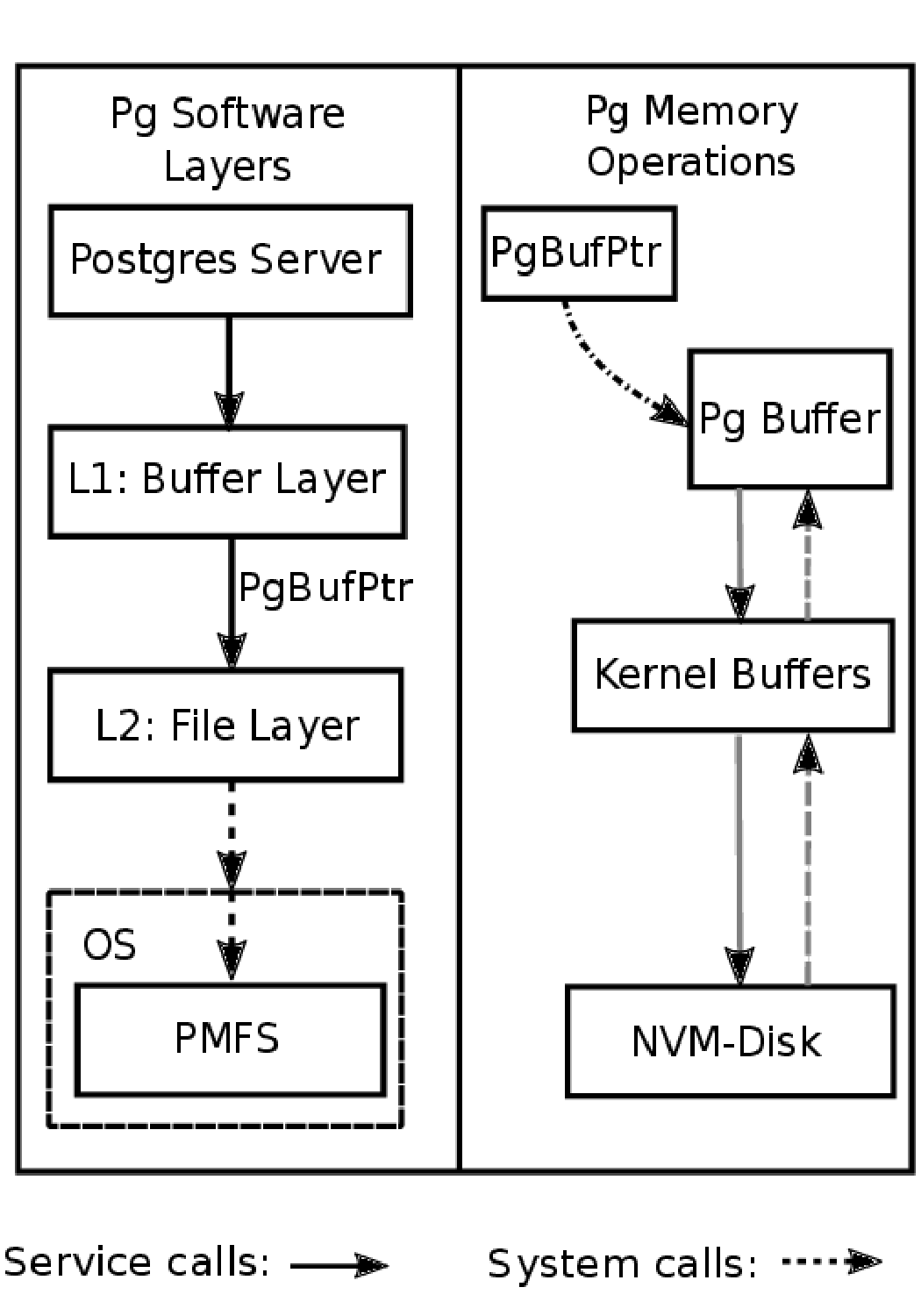}}
\subfigure[Modified storage engine - SE1]{\label{Fig3b}\includegraphics[width=38mm]{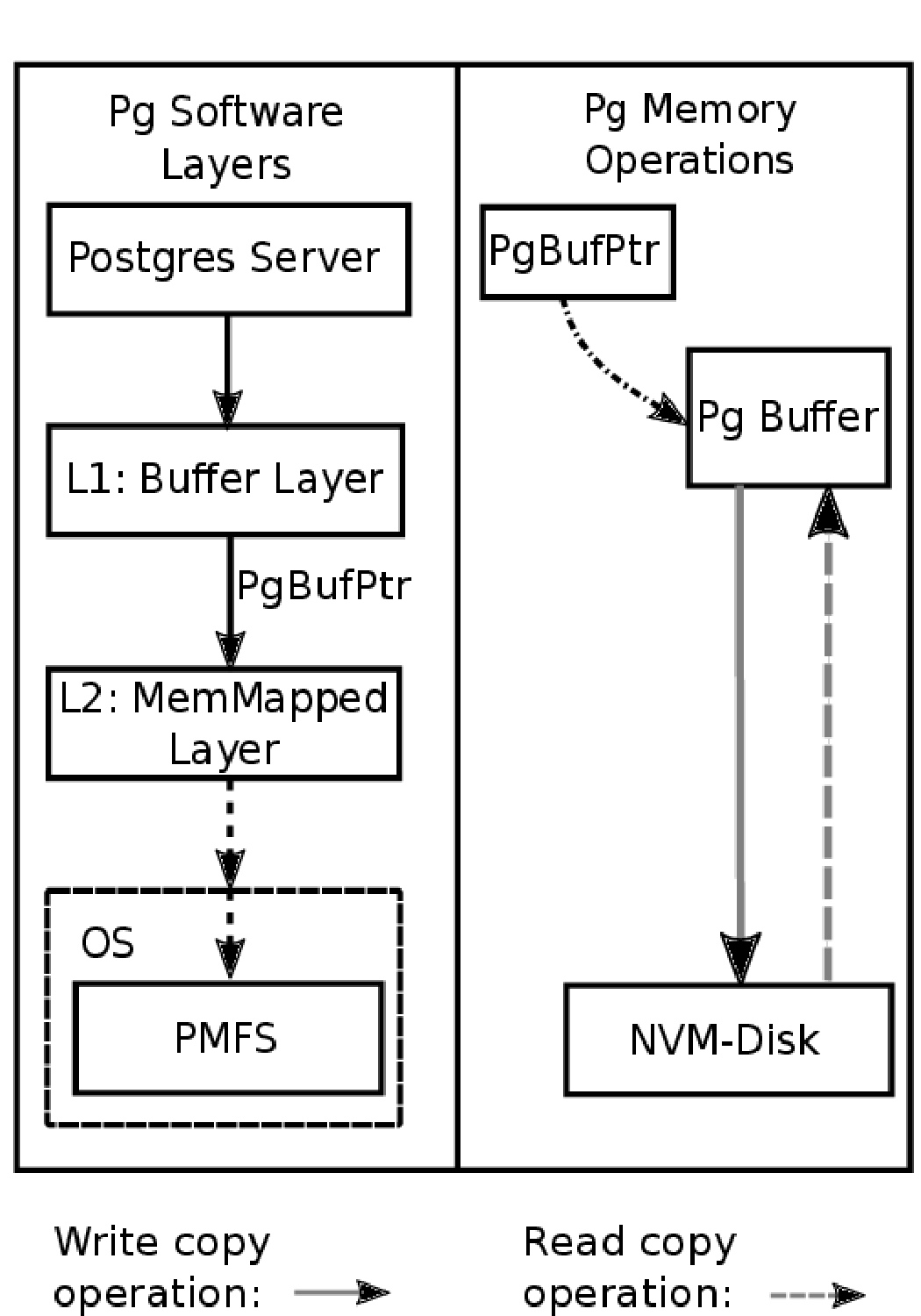}}
\subfigure[Modified storage engine - SE2]{\label{Fig3c}\includegraphics[width=38mm]{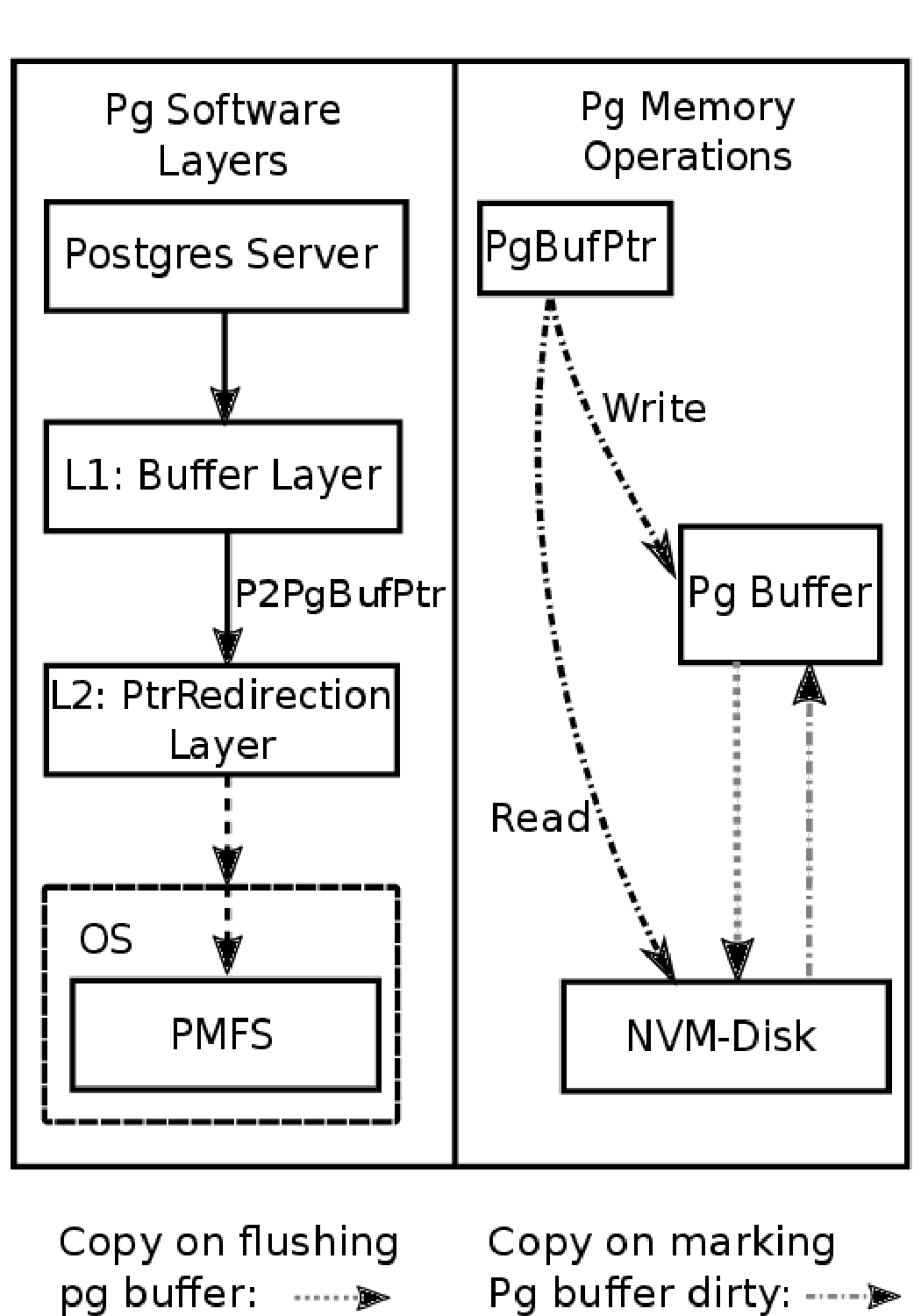}}
\caption{High level view of read and write memory operations in PostgreSQL (read as ``pg'' in short form) and modified SEs}
\label{Fig3}
\end{figure*}

\section{A Case Study: P\MakeLowercase{ostgre}SQL}
\label{sec:CaseStudy}
\noindent PostgreSQL is an open source object-relational database system. It is fully atomicity, consistency, isolation, durability (ACID) 
compliant and runs on all major operating systems including Linux \cite{momjian2001postgresql}. In this section, we study the 
storage engine (SE) of PostgreSQL and apply necessary changes to make it NVM-aware. We first describe the read-write architecture 
of PostgreSQL and then explain our modifications.

\subsection{Read-Write Architecture of PostgreSQL}

Fig.~\ref{Fig3a} shows the original PostgreSQL architecture from the perspective of read and write file operations. 
The left column in the figure shows the operations performed by different software layers of PostgreSQL, while the right column 
shows the corresponding data movement activities. Note that in Fig.~\ref{Fig3a} we assume the disk has already been replaced 
by NVM hardware with PMFS as the file system. However, the same behavior would be expected using a regular disk and a traditional 
FS, since PostgreSQL heavily relies on file I/O for read and write operations and the file I/O APIs in PMFS are the same as those 
in a traditional FS.

The PostgreSQL server calls the services of the  \textit{Buffer Layer} which is responsible for maintaining an 
internal buffer cache. The buffer cache is used to keep a copy of the requested page which is read from the 
storage. Copies are kept in the cache as long as they are needed. If there is no free
slot available for a newly requested page then a replacement policy is used to select a victim. 
The victim is evicted from the buffer cache and if it is 
a dirty page, then it is also flushed back to the permanent storage.

Upon receiving a new request to read a page from storage, the \textit{Buffer Layer} finds a free buffer cache slot 
and gets a pointer to it. The free buffer slot and corresponding pointer are shown in Fig.~\ref{Fig3a} as \textit{Pg Buffer} and \textit{PgBufPtr}, respectively. The \textit{Buffer Layer} then passes the pointer to the \textit{File Layer}. Eventually, the \textit{File Layer} of PostgreSQL invokes the file read and write system calls implemented by the 
underlying FS. 

For a read operation, PMFS copies the data block from NVM to a kernel buffer and then the kernel copies the requested 
data block to an internal buffer slot pointed by \textit{PgBufPtr}. In the same way, two copies are made for write operation but in the opposite direction.

Hence, the SE of original PostgreSQL incurs two copy operations for each miss in the internal buffer cache. This is 
likely to become a big overhead for databases running queries on large datasets. However, since PMFS can map the entire NVM address space into the kernel's virtual address space~\cite{dulloor2014system}, the copy overhead can be avoided by making modifications in the SE. We apply these modifications in two incremental steps as described in 
the following subsections.

\subsection{SE1: Using Memory Mapped I/O}
In the first step towards leveraging the features of NVM, we replace the \textit{File Layer} of PostgreSQL by a 
new layer named \textit{MemMapped Layer}. As shown in Fig. \ref{Fig3b}, this layer still receives a pointer 
to a free buffer slot from the \textit{Buffer Layer}, but instead of using the file I/O interface, it uses the memory 
mapped I/O interface of PMFS. We term this storage engine \textit{SE1}.

\noindent\textbf{Read Operation:} When accessing a file for a read operation, we first open the file using the \verb+open()+ 
system call, same as in original PostgreSQL. Additionally, we create a mapping of the file using \verb+mmap()+. 
Since we are using PMFS, \verb+mmap()+ returns a pointer to the mapping of the file stored in NVM. The implementation of \verb+mmap()+ by PMFS provides the application with direct access to mapped pages of files residing in NVM.

As a result, we do not need to make an intermediate copy of the requested page from NVM into kernel buffers. We can 
directly copy the requested page into internal buffers of PostgreSQL by using an implicit \verb+memcpy()+ as shown in Fig. \ref{Fig3b}. When all requested operations on a given file are completed and it is not needed anymore, 
the file can be closed. Upon closing a file, we delete the mapping of the file by calling the \verb+munmap()+ function. 

\noindent\textbf{Write Operation:} The same approach as in the read operation is used for writing data into a file. The file to be modified is first opened and a mapping is created using \verb+mmap()+. The data to be written into the file is copied 
directly from internal buffers of PostgreSQL into NVM using \verb+memcpy()+.

An SE with the above-mentioned modifications does not create an intermediate copy of the data in kernel buffers. Hence 
we reduce the overhead to one copy operation for each miss in the internal buffer cache of PostgreSQL. 

\subsection{SE2: Direct Access to Mapped Files}
In the second step of modifications to the SE, we replace the \textit{MemMapped Layer} of SE1 by 
the \textit{PtrRedirection Layer} as shown in Fig.~\ref{Fig3c}. Unlike the \textit{MemMapped Layer}, 
the \textit{PtrRedirection Layer} in SE2 receives the pointer to \textit{PgBufPtr} (i.e \textit{P2PgBufPtr}), which itself points to a free slot of the buffer cache. In other words, \textit{PtrRedirection Layer} receives a pointer to a pointer from the \textit{Buffer Layer}.

\noindent\textbf{Read Operation:} When accessing a file for a read operation, we first open the file using \verb+open()+ system call, same as in original PostgreSQL and SE1. Additionally, we also create a mapping of the file using \verb+mmap()+. Originally \textit{PgBufPtr} points to a free slot in the 
internal buffer cache. Since \verb+mmap()+ makes the NVM mapped address space visible to the calling process, the \textit{PtrRedirection Layer} simply redirects the \textit{PgBufPtr} to point to the corresponding address of the file residing in NVM. Pointer redirection in case of a read operation is shown by a black dashed arrow with the ``Read'' label in Fig.~\ref{Fig3c}.

As a result of pointer redirection and the visibility of the NVM address space enabled by PMFS, we incur no copy 
overhead for read operations. This can represent a significant improvement since read operations are predominant in queries that operate on large datasets.

\noindent\textbf{Write Operation:} PMFS provides direct write access for files residing in NVM. 
However, it does not manage the data consistency in memory mapped operations and leaves the responsibility 
to the application \cite{dulloor2014system} - i.e., PostgreSQL.

PostgreSQL is an ACID compliant DBMS which uses 
multi-version concurrency control (MVCC) \cite{neumann2015fast}
to maintain data consistency.
Under MVCC, concurrent executing transactions see a snapshot of the data at a particular instant in time, 
regardless of the current state of the underlying data. This provides data consistency and transaction isolation \cite{PostgreSQLDocBook}. 

To keep the consistency model of PostgreSQL unaltered and functionally correct,
SE2 performs two actions before modifying the 
actual content of the page and marking it as dirty. First, if the page is residing in NVM, it copies the page back from NVM into the 
corresponding slot of the internal buffer cache, i.e. \textit{Pg-Buffer}. Second, it undoes the redirection of \textit{PgBufPtr} such that it 
again points to the corresponding slot in the buffer cache and not to the NVM mapped file. This is shown by a black dashed arrow with 
the ``Write'' label in Fig.~\ref{Fig3c}. This way, SE2 ensures that each transaction (or query) updates only its local copy of the page.

In other words, SE1 and SE2 always use the internal PostgreSQL  buffers for write operations, avoiding writes directly into  database files residing in NVM disk. As a result,  data consistency model of PostgreSQL is not violated and transactions are protected from viewing inconsistent data. In summary, SE1 and SE2 operate in the same way as far as write operation is concerned. However, for read operations, SE1 reduces overhead for each miss in internal buffer cache from two copy operations in original PostgreSQL storage engine to one. On the other hand, SE2 reduces this overhead to zero copy operations.

\subsection{Discussion}

Researchers have developed libraries to assist programmers in developing applications targeting NVM. One such collection of libraries is the Persistent Memory Development Kit (PMDK) \cite{PMDKLib} from Intel. It allows an application to directly access NVM resident data as memory mapped files through the DAX feature present in App Direct mode of Intel's 3D XPoint DIMMs (see Section~\ref{NVDIMM}).

The modifications shown in this section can be implemented with such a library when targeting a system with 3D XPoint technology, and would in essence perform the same operations our implementation does. For example, storage engine SE2 could benefit from the \textit{pmemobj} library of PMDK in order to mmap opened files, by using the API calls provided by the library. However, as we have done, it also requires similar modifications in the source code of PostgreSQL storage engine at appropriate places to embed the API calls. The advantage of using tools like PMDK is that modifications are based on standardized and tested API calls.
\section{Methodology}
\label{sec:methodology}

\noindent System-level evaluation for NVM technologies is challenging due to limited availability of real hardware. 
Software simulation infrastructures are a good fit to evaluate systems in which NVM is used as a DRAM replacement,
or in conjunction with DRAM as a hybrid memory system. However, when using NVM as a permanent storage replacement, 
most software simulators fail to capture the details of the operating system, and comparisons against traditional disks are not 
possible due to the lack of proper simulation models for such devices. As the authors of PMFS \cite{dulloor2014system} noted, 
an emulation platform is the best way to evaluate such a scenario.

\doublerulesep 0.1pt
\begin{table}[h]
  \caption{Test machine characteristics}
  \label{tab:machine}
  \begin{tabular}{@{}ll@{}}
  \toprule
  \textbf{Component} & \textbf{Description} \\ 
  \midrule
  \multirow{2}{*}{Processor}    & Intel Xeon E5-2670 @ 2.60Ghz \\
                                & HT and TurboBoost disabled \\
  \multirow{3}{*}{Caches}       & Private: L1 32KB 4-way split I/D, \\
								& L2 256KB 8-way \\
                                & Shared: L3 20MB 16-way \\
  Memory                        & 256GB DDR3-1600, 4 channels, delivering\\
								& up to 51.5GB/s\\ 
  OS                            & Linux Kernel 3.11.0 with PMFS support\\  &\cite{githubPMFS,dulloor2014system} \\
  \multirow{2}{*}{Disk storage} & Intel DC S3700 Series, 400GB, SATA 6Gb/s \\ 
                                & Read 500MBs/75k iops, Write 460MBs/36k iops\\
  PMFS storage                  & 224 GB of total DRAM \\
  \bottomrule
  \end{tabular}
\end{table}

\subsection{Emulation Platform}

We set up an infrastructure similar to that used by the PMFS authors. We first recompile the Linux kernel of our 
test machine with PMFS support. Using the \textit{memmap} kernel command line option we reserve a physically contiguous area of the 
available DRAM at boot-time, which is later used to mount the PMFS partition. In other words, a portion of the DRAM holds the disk 
partition managed by PMFS and provides features similar to those of NVM, such as byte-addressability and lower latency compared to a disk. 
Table~\ref{tab:machine} lists the test machine characteristics. We configure the machine to have a 224GB PMFS partition, leaving 32GB of 
DRAM for normal main memory operation. A high-end SSD is used as regular disk storage.

To fairly evaluate storage engines SE1 and SE2, we compare their performance with two baselines using unmodified PostgreSQL. A similar comparison approach is also adopted by other closely related and complementary works \cite{gao2011pcmlogging,son2017log} using NVM in context of disk-based DBMS as explained in Section~\ref{sec:RelatedWork}.
While comparing with prior work that employs NVM in context of in-memory
DBMS or in from-scratch NVM-aware DBMS designs is out of the scope of this paper, as these systems present different sets of features and target different domains, we do include a qualitative comparison in Section~\ref{sec:RelatedWork}.

The two baselines use unmodified  PostgreSQL 9.5 with the dataset stored in: (i) a regular high-end disk (\textit{disk\_base95}), and (ii) in the PMFS partition (\textit{pmfs\_base95}). The modified storage engines - SE1 and SE2 - are run with the dataset stored on the PMFS partition and are termed \textit{pmfs\_se1} and \textit{pmfs\_se2}, respectively. All the evaluated setups have 32GB of DRAM available, and we configure PotgreSQL with the parameters detailed in Table~\ref{ParamTable}.

\begin{table}[h]
  \caption{PostgreSQL configuration parameters}
  \label{ParamTable}
  \begin{tabular}{@{}ll@{}}
  \toprule
  \textbf{Parameter} & \textbf{Value} \\ 
  \midrule
  
  max\_wal\_size                        				& 2GB\\
  work\_mem                             				& 3.2GB\\
  effective\_cache								& 24GB\\
  shared\_buffers (i.e., buffer pool size)		& 8GB\\
  maintainance\_ work\_mem					& 32MB\\
   
  \bottomrule
  \end{tabular}
\end{table}


Since DRAM read latencies are expected to be similar to projected NVM read latencies \cite{mittal2016survey,arulraj2015let,wang2013low,chang2012limits}, the emulation platform we employed provides good performance estimations. Note that this system resembles a setup featuring 3D XPoint DIMMs in App Direct mode. In our experiments, we report wall-clock query execution times as well as data obtained with performance counters using the \textit{perf} toolset.

\subsection{Workloads}

\begin{figure*}
\centering
\includegraphics[width=\linewidth]{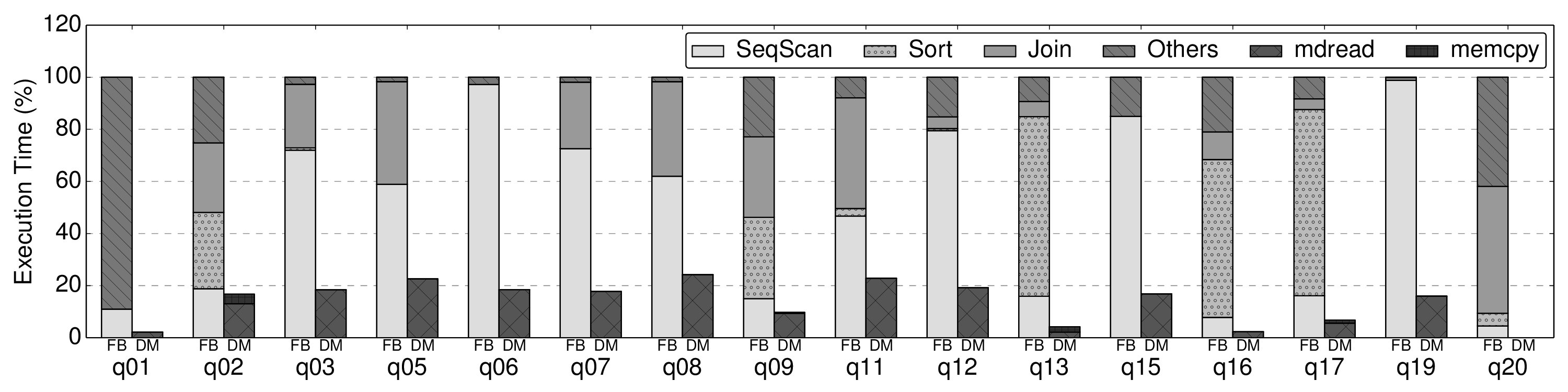}
\caption{Execution time breakdown for TPCH queries in traditional DBMS with database stored in disk-storage}
\label{query-breakdown}
\end{figure*}

TPC-H \cite{council2008tpc} is a widely used benchmark and a good representative of decision support system
(DSS) queries. Therefore, to evaluate our proposed SEs, we employ DSS queries from the TPC-H benchmark configured with a scale factor of 100, which leads to a dataset larger than 150GB when adding the appropriate
indexes. We build indices only on permitted columns of tables (i.e. primary or foreign keys) as specified in clause 1.5.4 of TPC-H specifications.

Like most data-intensive workloads, these queries are read dominant, which will enable us to draw accurate results from our emulation platform. We report results for 16 of the 22 TPC-H queries since some queries fail to complete under PMFS storage, even when executed with the unmodified PostgreSQL storage engine (i.e. baseline \textit{pmfs\_base95}).

Fig.~\ref{query-breakdown} shows the characterization of the different TPC-H queries in the form of an execution 
time breakdown. The data is collected using a scale factor of 100 with a baseline system that uses a high-end disk as primary storage 
(\textit{disk\_base95}).  The  figure shows two bars: 
functional breakdown (FB) and data movement (DM). FB shows 
the percentage of execution 
time spent across the most relevant database operators, i.e., sequential scan (SeqScan), Sort, Join and all other operations
combined together. DM shows the percentage of execution time spent in the main function performing data reads from 
disk (\textit{mdread}) and also \textit{memcpy} since it is used internally by the kernel to bring data into the application buffers.

As can be seen in the figure, most of the queries are dominated
by sequential scan operations, as expected from read-dominant queries.
This is confirmed by the fact
that most queries spend about 20\% of their execution time bringing
data in from storage to application-level buffers, as shown by
the DM bar.  These overheads are expected to become worse with larger datasets in the future, therefore
lowering the data access latency and avoiding unnecessary data movement is
critical to reduce query execution time. Note that these overheads are due to
data movement operations that can be avoided by reading directly from primary storage with our proposed NVM-aware SEs.

\section{Evaluation}
\label{sec:evaluation}
\noindent In this section, we show the performance impact of the modified storage engines (SE) on kernel execution time and on wall-clock execution time for TPC-H queries. Later, we identify potential issues current DBMSs and applications, in general, may face in order to harness the benefits of  directly accessing data stored in NVM memory.

\subsection{Performance Impact on Kernel Execution Time}

\begin{figure*}
\centering
\includegraphics[width=\linewidth]{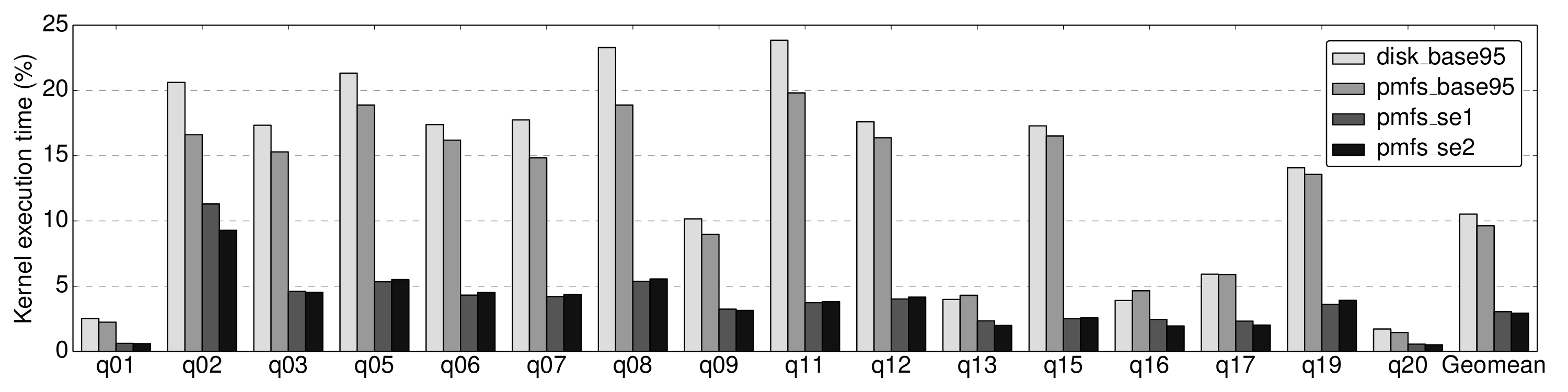}
\caption{Percentage of kernel execution time for each query}
\label{fig:kernel-time}
\end{figure*}

Fig.~\ref{fig:kernel-time} shows the percentage of kernel execution time (KET) for each of the evaluated queries 
running on the four evaluated 
systems. When using traditional file operations (e.g. \verb+read()+), like those employed in unmodified PostgreSQL, the bulk of the work when 
accessing and reading data is done inside the kernel. As can be seen, the baseline systems spend a significant amount of the execution time in 
kernel space: up to 
23.85\% 
(Q11 - \textit{disk\_base95}) and 
19.80\% 
(Q11 - \textit{pmfs\_base95}), with an average of around 10\%. 
KET is dominated by 
the time it takes to fetch data from the storage medium into a user-level buffer. These overheads are high in both disk and NVM storage, 
and are likely to increase as datasets grow in size.

However, when using \textit{SE1} or \textit{SE2}, this data movement can be minimized or even avoided. 
For \textit{pmfs\_se1} we observe that the amount of time spent in kernel space decreases substantially 
and it is very similar to that observed for \textit{pmfs\_se2}. This is because the two systems are 
doing a similar amount of work on the kernel side, with the difference that \textit{SE1} is doing an 
implicit \verb+memcpy()+ operation into a user-level buffer. 
Overall, we see that the modified SEs are able to reduce KET significantly in  most queries: 
Q02 to Q12, Q15, and Q19. A few queries show lower reductions because they operate over a small amount of 
data, e.g, Q01, Q13, Q16, and Q20. 
 
 Reduction in KET is consistent with the 
 data movement (DM) bar in Fig. \ref{query-breakdown}. 
 For example, queries with relatively more time spent in DM operations 
 (e.g., Q02 to Q12, Q15, and Q19) show a higher reduction in KET  
 when executed using SE1 and SE2, as shown in Fig. ~\ref{fig:kernel-time}. On the other hand, 
 queries that spend relatively lesser time in DM operations (e.g., Q01, Q13, Q16, and Q20) show a  lower reduction in KET. 
 In other words, read intensive  queries show more reduction in KET 
 when executed using SE1 and SE2.
 An important point to note is that for \textit{SE1} and \textit{SE2} the kernel space time is likely to remain near constant as datasets grow since no work is done to fetch data.

\subsection{Query Performance Improvement}

\begin{figure*}
\centering
\includegraphics[width=\linewidth]{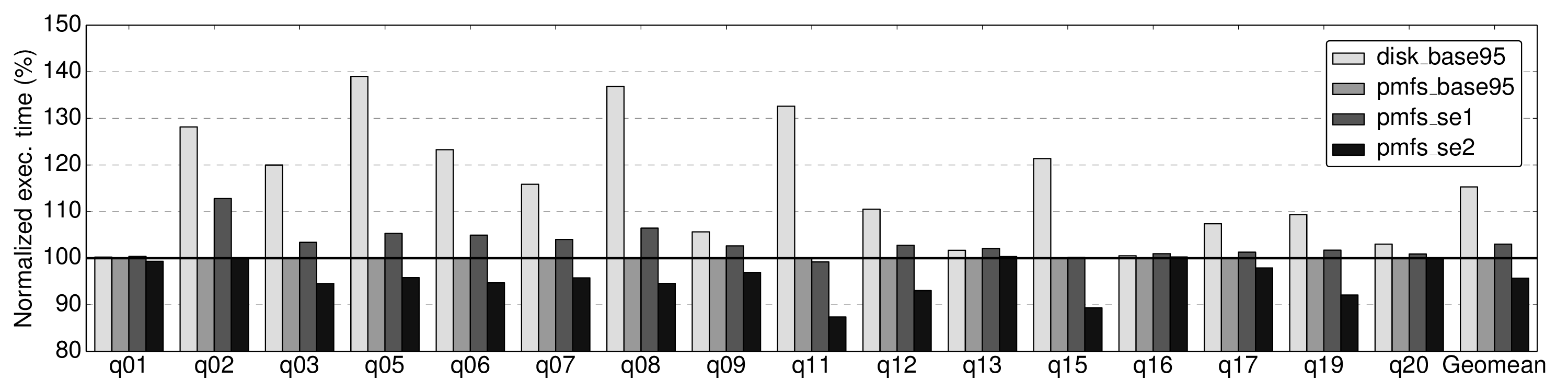}
\caption{Wall-clock execution time normalized with respect to \textit{pmfs\_base95}}
\label{fig:exec-time}
\end{figure*}

Fig.~\ref{fig:exec-time} shows wall-clock execution time for each query on the evaluated systems. The data is 
normalized with respect to \textit{pmfs\_base95}. We observe that the benefits of moving from disk to a faster storage can be high for read-intensive queries such as Q05 (39\%), Q08 (37\%), and Q11 (33\%). 
However, for compute-intensive queries, such as Q01 and Q16, 
the benefits are non-existent. On average, the overhead of using disk over PMFS storage is around 15\%.

For \textit{SE1}, the reductions observed in terms of kernel execution time do not translate into reductions in overall query execution 
time. The main reason for this is the additional \verb+memcpy()+ operation performed to copy the data into the application buffer. In fact, 
we find that this operation in PMFS is sometimes slower than the original \verb+read()+ system call employed in the baseline, leading to a 
3\% slowdown on average.

When using \textit{SE2} there is no data movement at the time of fetching data into an application-accessible memory region, due 
to the possibility of directly referencing data stored in PMFS. However, this has a negative side effect when accessing the data 
for processing later on, as it has not been cached by the processing units. Therefore, the benefits of avoiding data movement to 
make it accessible are offset by the penalty to fetch this data close to the processing units at a later stage. 
In order to mitigate this penalty, \textit{SE2} incorporates a simple software prefetching scheme that tries to fetch in advance 
the next element to be processed within a data block. 

When compared to \textit{pmfs\_base95}, \textit{SE2} is able to achieve significant performance improvements in read-dominant queries such as Q11 (13\%), Q15 (11\%), and Q19 (8\%). These performance improvements are also consistent with the data presented in Fig.~\ref{query-breakdown}. Queries that are dominated by sequential scan operations are the ones that benefit from our modified storage engines. For example, although Q02 spends almost 18\% of its execution time in data movement operations (like Q15 and Q19, as shown by the DM bar in Fig. \ref{query-breakdown}), sequential scan makes only 20\% of the database operations performed by the query (unlike Q15 (84\%) and Q19 (98\%)) as shown by the FB bar in the same figure. As a result, Q02 does not benefit from our modified storage engine. The same explanation is valid for Q13, Q16, Q17, and Q20. On average, \textit{SE2} is around 4\% faster than \textit{pmfs\_base95} and around 19\% faster than \textit{disk\_base95}.

\begin{figure*}
\centering
\includegraphics[width=\linewidth]{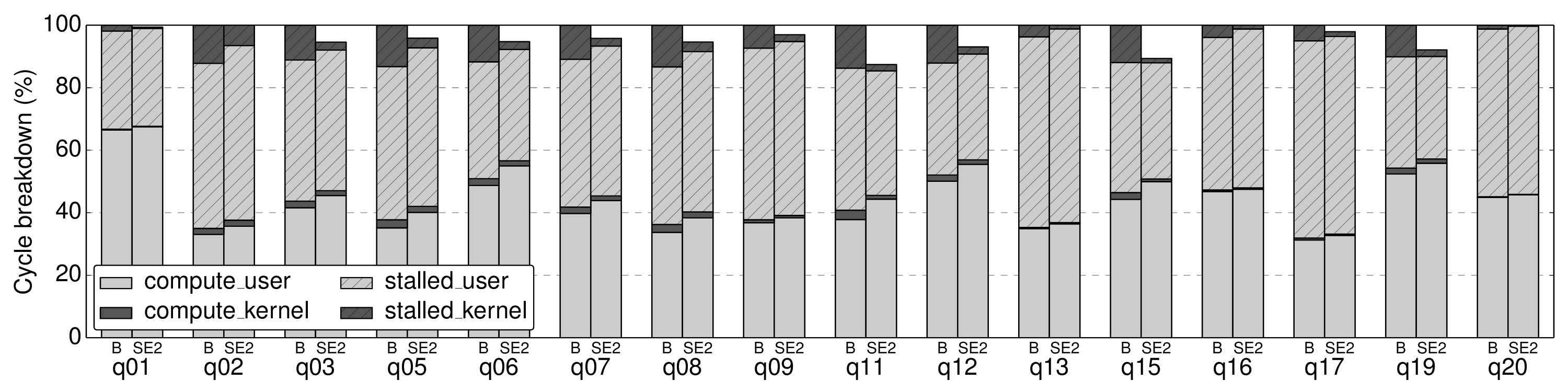}
\caption{Execution-time breakdown for compute and stalled cycles --- B $=$ \textit{pmfs\_base95}, SE2 $=$ \textit{pmfs\_se2}}
\label{fig:compute-stall}
\end{figure*}

\begin{figure*}
\centering
\includegraphics[width=\linewidth]{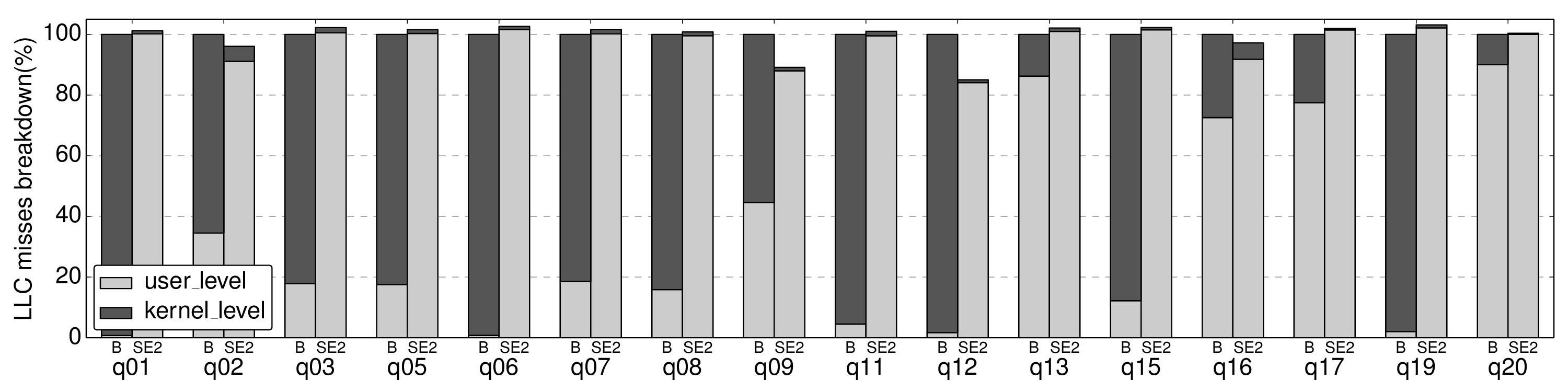}
\caption{Last-level cache (LLC) misses breakdown --- B $=$ \textit{pmfs\_base95}, SE2 $=$ \textit{pmfs\_se2}}
\label{fig:llc-misses}
\end{figure*}

Fig.~\ref{fig:compute-stall} shows a classification of each cycle of execution as {\lq{compute}\rq}, if at least one 
instruction was committed during that cycle, or as {\lq{stalled{\rq} otherwise. These categories are further broken down 
into user and kernel level cycles. Data is shown for \textit{pmfs\_base95} and \textit{SE2}, normalized to the former. 
As can be seen, the \textit{stalled\_kernel} component correlates well with the kernel execution time shown in Fig.~\ref{fig:kernel-time}, 
and this is the component that is reduced in \textit{SE2} executions. 
Furthermore, reductions in the \textit{stalled\_kernel} component using SE2 are proportional to the 
time spent in DM operations shown in Fig.~\ref{query-breakdown}. 
We observe that for most queries some of the savings from \textit{stalled\_kernel} 
shift to \textit{stalled\_user} since data needs to be brought close to the processing unit when it is needed for processing. 
There are some exceptions, i.e., Q11, Q15, and Q19, for which the simple prefetching scheme is able to mitigate this fact effectively.

Fig.~\ref{fig:llc-misses} shows a breakdown of user and kernel last-level cache (LLC) misses. Here, 
we can clearly see how the number of LLC misses remains quite constant when comparing \textit{pmfs\_base95} and \textit{SE2}, 
but the misses shift from kernel level to user level. Moreover, in our experiments, we observe that user level misses have a 
more negative impact in terms of performance because they happen when the data is actually needed for processing, and a full LLC 
miss penalty is paid for each data element. On the other hand, when moving larger data blocks to an application buffer, optimized 
functions are employed and the LLC miss penalties can be overlapped.

\subsection{Discussion}~\label{Discussion}
Experimental results show that there is a mismatch between the potential performance benefits shown in Fig.~\ref{fig:kernel-time} and the actual benefits 
in terms of wall-clock query execution time 
shown in Fig.~\ref{fig:exec-time}. Direct access to memory regions holding persistent data can provide 
significant benefits, but this data needs to be close to the processing units when it is needed. To this end, we employ 
simple software prefetching schemes that provide moderate average performance gains using our SE2 engine. However, carefully crafted ad-hoc software 
prefetching is challenging, and applications may not be designed in a way that makes it easy to hide long access latencies even 
with the use of prefetching, as happens with PostgreSQL. Moreover, such a solution is application and architecture dependent.

Additional software libraries and tools that aid programmability are needed in such systems. Such libraries could implement solutions 
like helper threads for prefetching particular data 
regions, effectively bringing data closer to the core (e.g., LLC) with minimal application interference. This approach would 
provide generic solutions for writing software that takes full advantage of the capabilities that NVM can offer.

Therefore, to further improve the performance when using our SE2 engine, we develop a general purpose data prefetching library employing helper threads to prefetch data while the main computational thread continues to make forward progress. By exposing a simple API to the programmer, a user can specify the desired number of threads to use, their mapping to cores, and enqueue jobs to prefetch desired data regions.

\section{Software library for NVM data prefetching}
\label{sec:library}
\noindent The data readiness problem is likely to appear in any application that directly accesses large pools of data residing in NVM. In this section, we provide details on our general purpose data prefetching library, that aims to aid programmability to easily prefetch desired data regions. The library uses POSIX threads and implements a simple API to create, control and assign jobs to threads.

\subsection{Helper Threads}

Prefetching data~\cite{annavaram2001data} reduces the cache miss rate and hence
accelerates an application\textquotesingle s execution. An in-advance knowledge of memory regions
to be accessed can be used to prefetch data into caches before it is needed. However,
the application should not be stalled while prefetching the data. This can be achieved
by using independent helper threads for data prefetching.

Synchronization between the main computation thread and helper thread(s) is 
important~\cite{jung2006helper}. Prefetching data too early before it is needed
by the computation thread can result in cache pollution. Furthermore, required
cache lines may get evicted before they are accessed. Similarly, prefetching data
too late is also not useful, rather counter-productive. It can also lead to cache
pollution and degrade performance.

In our implementation, a helper thread is a simple block of code. Given a starting
memory address and the amount of data to be prefetched, the helper thread prefetches
data into caches without interfering with the main computation thread. We employ a
job queue to build a single producer - (multiple) consumer relationship between
a computation thread and one or multiple helper threads. We employ light-weight 
compare and swap instructions for synchronization in the job queue between a 
computation thread and the different helper threads.

A computation thread places the starting address and the amount of data to be prefetched into a job unit and enqueues it into the job queue. On the other end, a helper thread picks the job item, unpacks it and then prefetches the data into caches. Data prefetching is performed without stalling the computation thread.

\subsection{Library Services}
A programmer is responsible for inserting API calls to construct helper threads. However, as SE2 already has knowledge of the size and location of the block to be read, inserting data prefetching APIs in SE2 source code is not a tedious task. Our library provides three basic services via a simple API.

\begin{enumerate}
 \item \textbf{Creation of helper threads:} The library supports the  creation of a user-specified number of helper threads per computation thread, that are synchronized using a job queue. A slightly different thread creation policy is also supported, as explained in detail in Section 7.3.
 \item \textbf{Assigning work to helper threads:} Work is assigned to helper threads by placing jobs into their job queue. On arrival of a job into the queue, helper threads wake up, one fetches the job from the queue and starts the data prefetching. After completing the job, the helper thread again waits for the next job\textquotesingle s arrival if the queue is empty.
 \item \textbf{Mapping threads to cores:} Our library also supports the selection of a core (in a multicore platform) on which a particular helper thread is to be executed by setting thread affinities. The affinities can also be set with respect to the computational thread, i.e., selecting the same or a different core.
\end{enumerate}

\subsection{Thread Mapping Schemes for Our Case Study}
\label{sec:mapping-schemes}

\noindent As discussed in Section~\ref{sec:evaluation}, \emph{SE2} directly accesses data located in NVM disk, without copying it into any local buffer. This direct access results in high cache miss rates when the data is needed for processing. However, due to ad-hoc data prefetching, SE2 achieves performance improvements for a few queries (i.e., Q11, Q15,  and Q19), but not for the majority of them. By performing ad-hoc data prefetching we were able to prefetch some blocks into caches, but not most of them due to the overheads it entailed performing the prefetching inside the computation thread. Furthermore, ad-hoc placement of data prefetching in the source code of an application can be tedious and difficult to maintain.

By using our data prefetching library services, data can be brought closer to processing cores before 
it is needed. When accessing a file for a read operation, \emph{SE2} creates a memory mapping of the file 
using \verb+mmap()+. Additionally, it has knowledge of the location and size of the data block to be read. Therefore, \emph{SE2} can pack this information into a job and place it into the job queue to be processed by helper threads. By using this approach, PostgreSQL can continue with its computation while the required data is prefetched into caches by a helper thread. 

Due to the way PostgreSQL is structured, it is not necessary to have more than one helper thread active to service a queued job in time before the next arrives. Therefore, we propose two mappings that employ only one helper thread and a third mapping that employs a different thread creation policy by instantiating two helper threads that work in tandem as we explain in the following subsections.

\begin{figure*} 
\centering     
\subfigure[\emph{M1} - Different physical core]{\label{CaseA}\includegraphics[width=32mm]{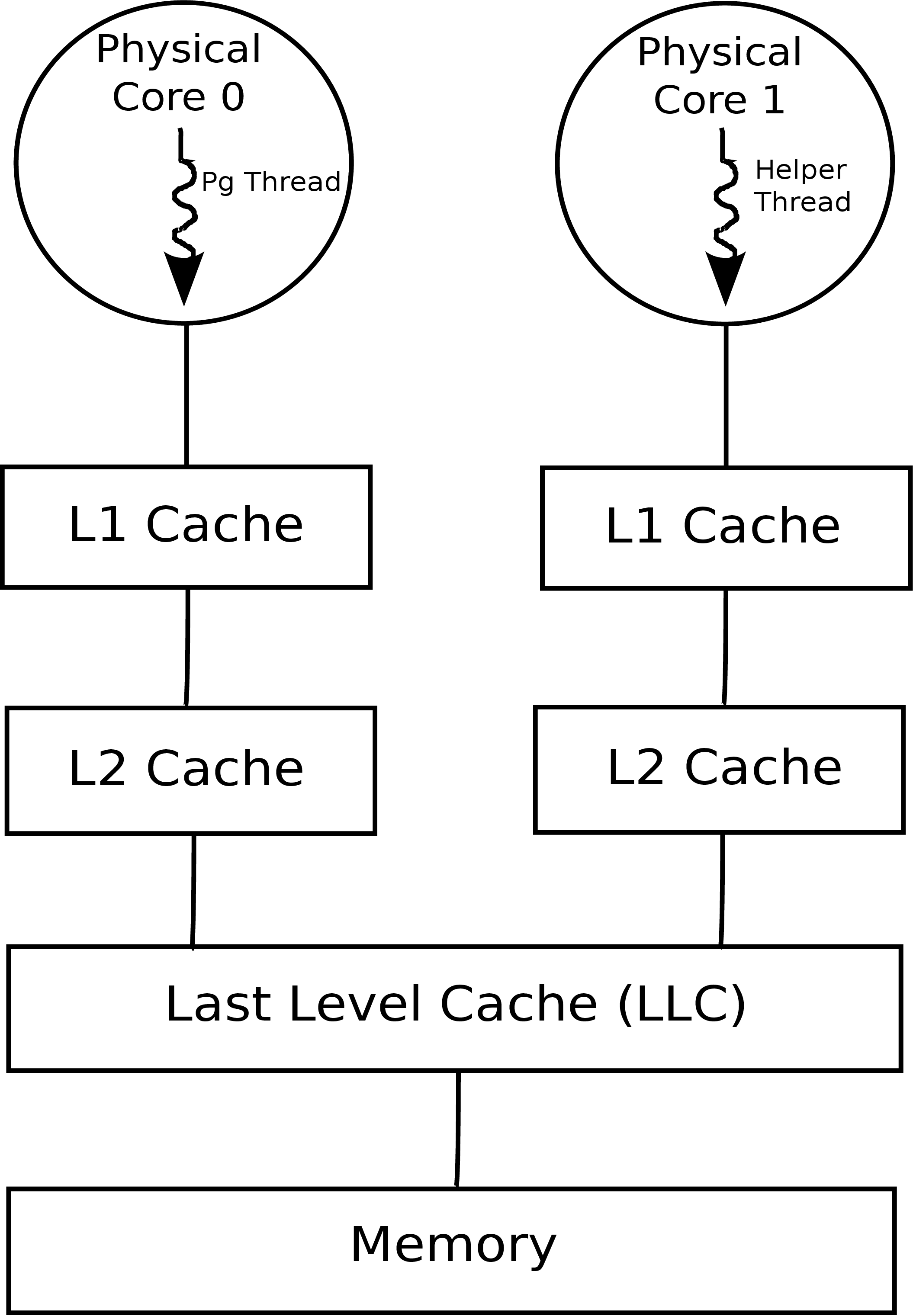}}
\subfigure[\emph{M2} - Same physical core (HT)]{\label{CaseB}\includegraphics[width=52mm]{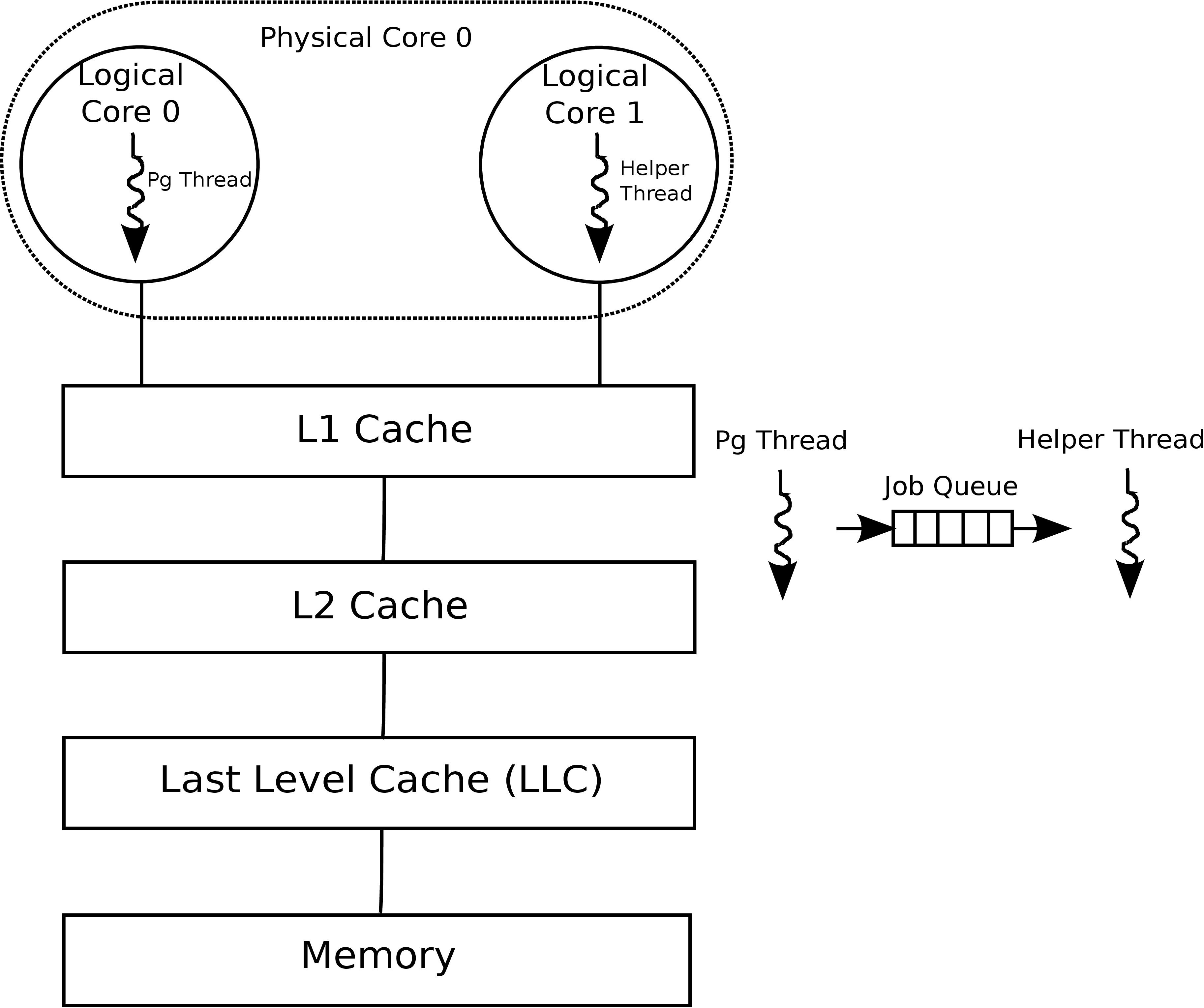}}
\subfigure[\emph{M3} - Thread mapping for the two helper threads scheme]{\label{CaseE}\includegraphics[width=87mm]{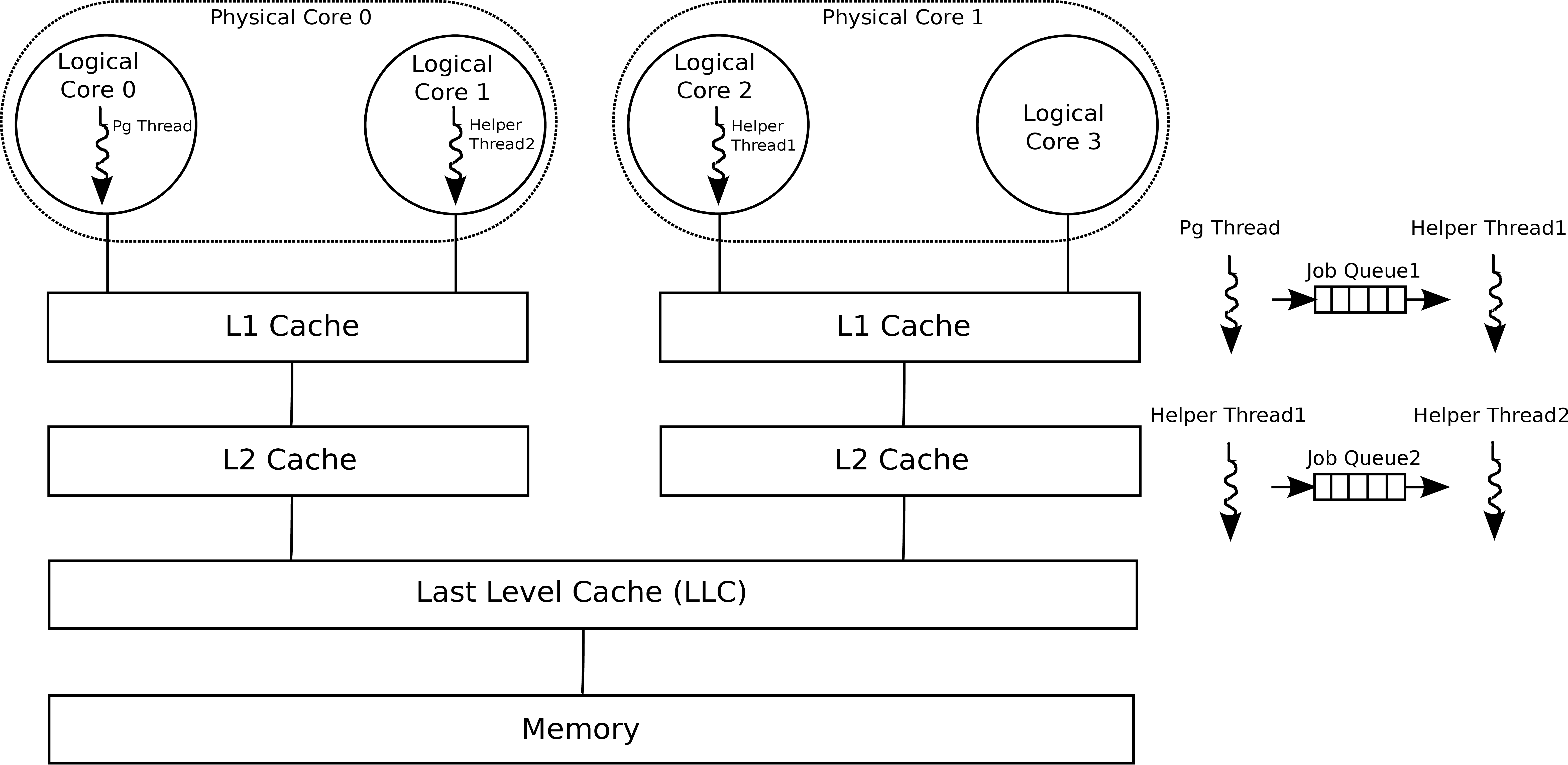}}
\caption{Different Helper thread mapping schemes with and without hyper-threading (HT) enabled}
\label{Thread-Mapping}
\end{figure*}

\subsubsection{Single helper thread}

When using a single helper thread there are two options for thread mapping. Mapping it to a different core than that of the computation thread, or to the same core. The latter option makes sense if the target machine supports hyper-threading - i.e., two hardware thread contexts per core. We explore these thread mappings:
\begin{enumerate}
 \item \verb+M1+ - Map helper thread to a different physical core, as shown in Fig.~\ref{CaseA}. In this case, each thread resides in a different core and hence prefetching is not done at the level of private caches but at the level of the last level cache, which is shared across cores.
 \item \verb+M2+ - Map helper thread into the same physical core while making use of hyper-threading, as shown in Fig.~\ref{CaseB}. Both threads reside within the same physical core, hence prefetching will also populate the private L1 and L2 caches present in the core.
\end{enumerate}

\subsubsection{Two helper threads}

As described for case \emph{M2}, the helper thread prefetches data into the private caches of the core where the computation thread is executing. However, in this scenario, the helper thread competes with the computation thread for hardware resources. This competition can slow down the execution of the computation thread. On the other hand, for case \emph{M1} there is no such competition for hardware resources at the expense of prefetching into the LLC, further away from the processing core.

A good compromise can be achieved using two helper threads working together. One helper thread is mapped to a different physical core than that of the computation thread and will process the jobs enqueued by the computation thread, similar to case \emph{M1}. Once this first thread finishes processing the job, it enqueues the same job into a second job queue that is processed by the second helper thread that is mapped into the same physical core as the computation thread. This thread mapping scheme which we term \verb+M3+ is shown in Fig.~\ref{CaseE}. The rationale behind this proposal is that the high penalty miss from main memory to LLC will be paid by a different core, while the helper thread residing on the same core as the computation thread will put data into the private caches and complete jobs much faster since data will be already present in the LLC.

\section{Evaluation of the data prefetching library}
\label{sec:library-evalualtion}

\begin{figure*}
\centering
\includegraphics[width=\linewidth]{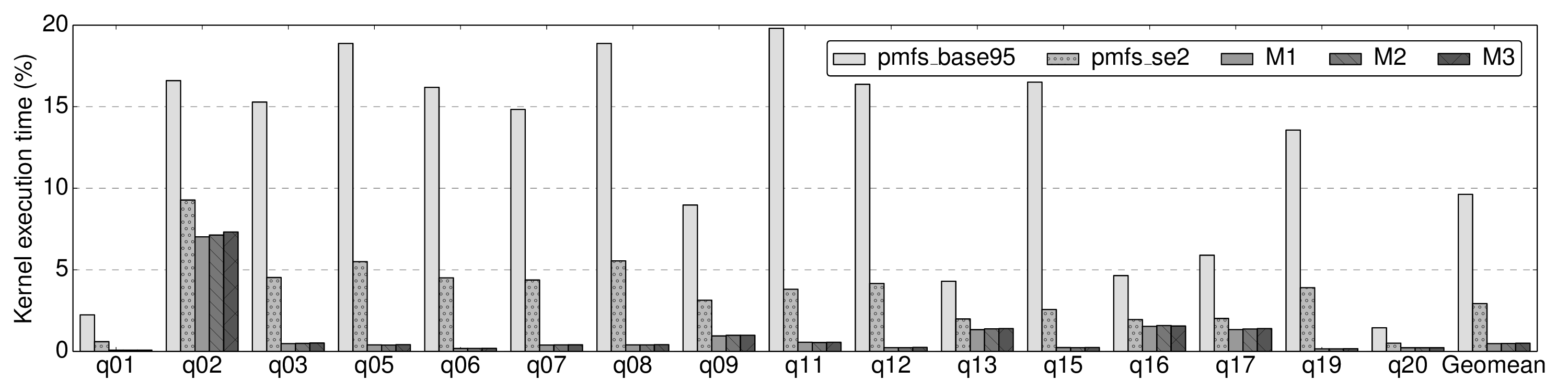}
\caption{Percentage of kernel execution time for PostgreSQL thread}
\label{kernel-time-pg}
\end{figure*}

\noindent In this section, we evaluate our modified storage engine \emph{SE2} while using the data prefetching library. 
For the scenarios in which we employ the library, we remove the simple ad-hoc software prefetching scheme used in the previous evaluation. However, the \emph{pmfs\_se2} system to which we compare does include the same ad-hoc prefetching used in the previous evaluation (Section~\ref{sec:evaluation}). The test machine and methodology employed is the same as explained in Section~\ref{sec:methodology}, with the exception that hyper-threading (HT) is enabled for \emph{M2} and \emph{M3}.

\subsection{Performance Impact on Kernel Execution Time}

Fig. \ref{kernel-time-pg} shows the percentage of kernel execution time of the PostgreSQL thread (computation thread) for each of the evaluated queries. As explained in Section~\ref{sec:evaluation}, \emph{SE2} only redirects the buffer pointer for file read operation from the local buffer cache to an NVM disk address that is within the address space of the PostgreSQL process. Hence, there is no data movement at the kernel level. As a result, the involvement of kernel in data movement and hence the average percentage of kernel execution time is already low in \emph{pmfs\_se2} as compared to \emph{pmfs\_base95}, reducing from 10\% to 3\%.

When using \verb+M1+, \verb+M2+, and \verb+M3+ helper thread schemes, by offloading the prefetching of entire blocks of data to helper threads, we can further hide kernel execution time overheads for the PostgreSQL thread running the query. Helper threads are more effective in prefetching data than the ad-hoc scheme used in pmfs\_se2. Therefore, the average percentage of kernel execution time further reduces from 3\% for \emph{pmfs\_se2} to 0.5\% in \verb+M1+, \verb+M2+, and \verb+M3+.

There is no noticeable difference between the three thread mapping schemes in terms of percentage of kernel execution time. The reason is that all three mapping schemes place data blocks at least into the LLC, which is enough to hide kernel related events such as page faults. 

\begin{figure*}
\centering
\includegraphics[width=\linewidth]{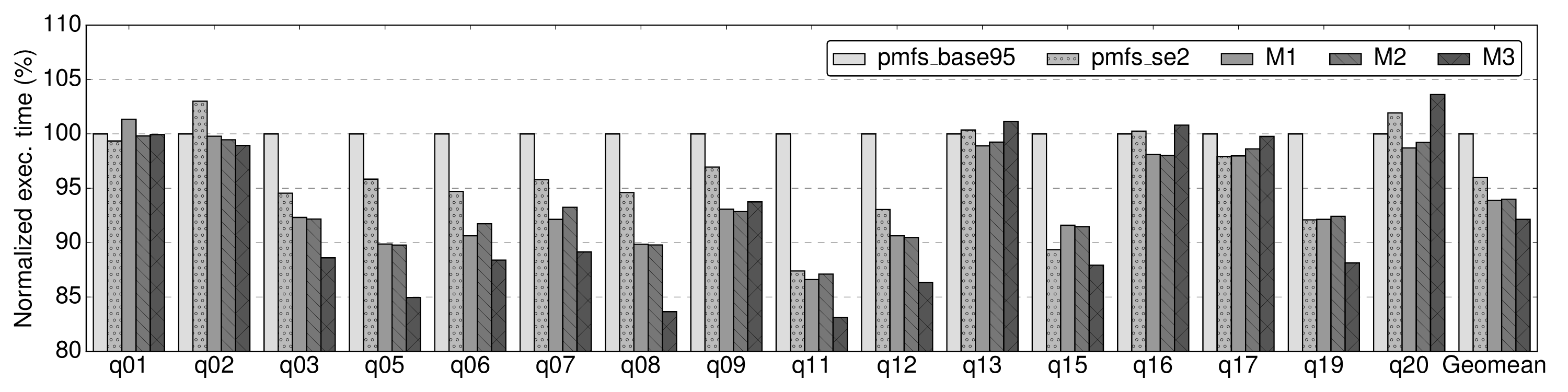}
\caption{Wall-clock execution time normalized with respect to pmfs\_base95}
\label{exec-time-new}
\end{figure*}
\begin{figure*}
\centering
\includegraphics[width=\linewidth]{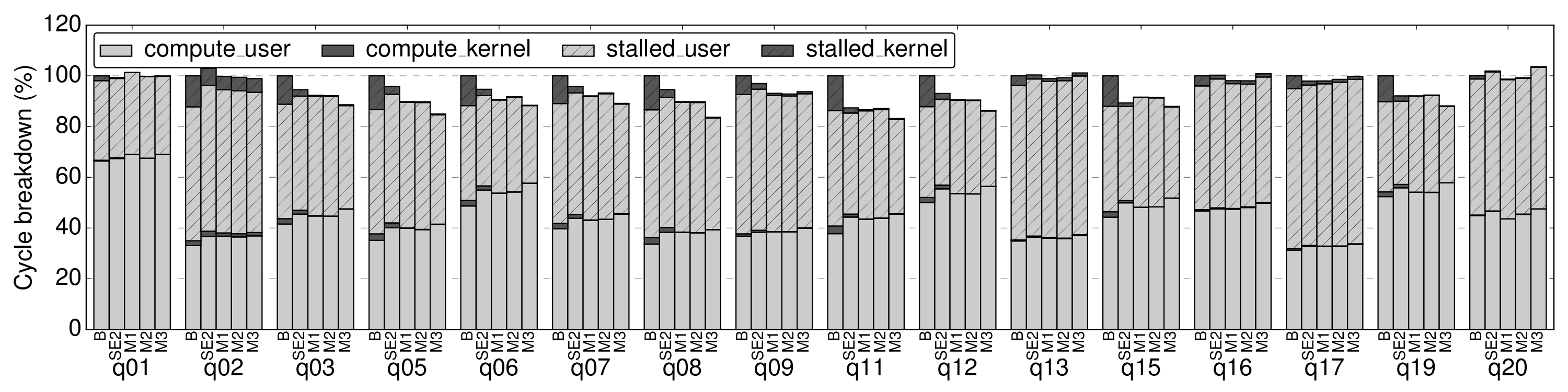}
\caption{Execution time breakdown into compute and stall cycles for PostgreSQL thread, normalized with respect to pmfs\_base95}
\label{compute-stall-pg}
\end{figure*}

\subsection{Query Performance Improvement}\label{QPI}

Fig.~\ref{exec-time-new} shows the wall-clock query execution time normalized with respect to \emph{pmfs\_base95} for all queries. We can observe that for queries in which \emph{pmfs\_se2} obtained better performance, the new evaluated systems with our prefetch library overall obtain better execution times, especially for the \verb+M3+ thread mapping scheme. M3 shows noticeable performance improvements for queries where the sequential scan operation represents a significant fraction of the total database operations - i.e. Q03-Q12, Q15, and Q19, as shown in Fig. \ref{query-breakdown}. On the other hand, queries where sequential scan operation consume less time (i.e. Q01, Q02, Q13, Q17, and Q20), show no performance improvement. On average, M3 obtains an 8\% performance improvement over the baseline. \verb+M1+ and \verb+M2+ show up to 13\% performance improvement (Q11), with an average of 6\% when compared to pmfs\_base95. 

Query execution time is mostly affected by two factors: cache misses and competition for hardware resources between threads mapped on the same physical core. Reducing any of these two factors should lead to better query execution times. To understand the improvements seen in Fig.~\ref{exec-time-new}, we provide insights in terms of compute and stalled core cycles and L1 cache misses for the PostgreSQL thread in Fig.~\ref{compute-stall-pg} and Fig.~\ref{L1-misses-pg}, respectively.

\begin{figure*}
\centering
\includegraphics[width=\linewidth]{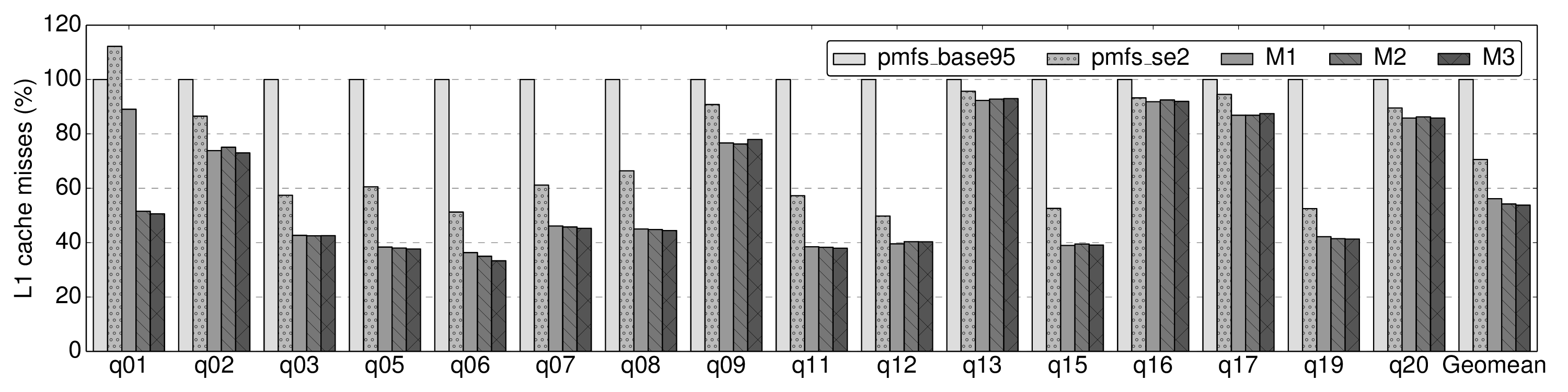}
\caption{L1 cache misses for PostgreSQL thread normalized with respect to pmfs base95}
\label{L1-misses-pg}
\end{figure*}

Fig.~\ref{compute-stall-pg} shows an execution cycle break down into the stall and compute cycles for all evaluated queries, but just for the PostgreSQL thread. An execution cycle is classified as `compute', if at least one instruction is committed during that cycle, or as `stalled' otherwise. Both \verb+M1+ and \verb+M2+ generate similar results in terms of wall-clock execution time. In Fig.~\ref{compute-stall-pg}, we can observe that both are able to reduce the kernel stall and compute components due to less kernel involvement in the main computation thread since the prefetching and kernel related events such as page faults are handled by the helper thread. However, the user level components are still very similar. This is because, as shown in Fig.~\ref{L1-misses-pg}, the actual number of L1 cache misses is also similar despite helper threads prefetching at different levels of the memory hierarchy. We attribute this to hardware prefetchers being much more efficient once the data is already in the LLC. M3 shows better performance than M1 and M2 as it maps the helper-thread, which handles the page faults, on a core different than that of compute thread as shown in Fig. \ref{CaseE}. Nonetheless, Fig.~\ref{L1-misses-pg} shows a large L1 cache misses reduction when compared to both \emph{pmfs\_base95} and \emph{pmfs\_se2}, proving that the prefetching library is performing well in hiding them from the main computation thread.

We can see in Fig.~\ref{compute-stall-pg} that \verb+M3+, besides being able to reduce the kernel components, also reduces the time spent in stalled user significantly, e.g., in Q03, Q05, Q06, Q12, and others. This is because of two factors: (i) prefetching into the L1 cache with our library is more timely than hardware prefetching, which might still be in-flight when the data is actually needed; and (ii) by using the two thread mapping approach we are able to reduce the overhead of the helper thread that is sharing the same physical core with the computation thread. Therefore, the first helper thread in \verb+M3+ brings the data into the LLC without interfering with the PostgreSQL thread that is running on a different physical core. While the second thread running on the same core brings the data into the private caches while incurring a lower overhead due to lower latency misses.

\begin{figure*}
\centering
\includegraphics[width=\linewidth]{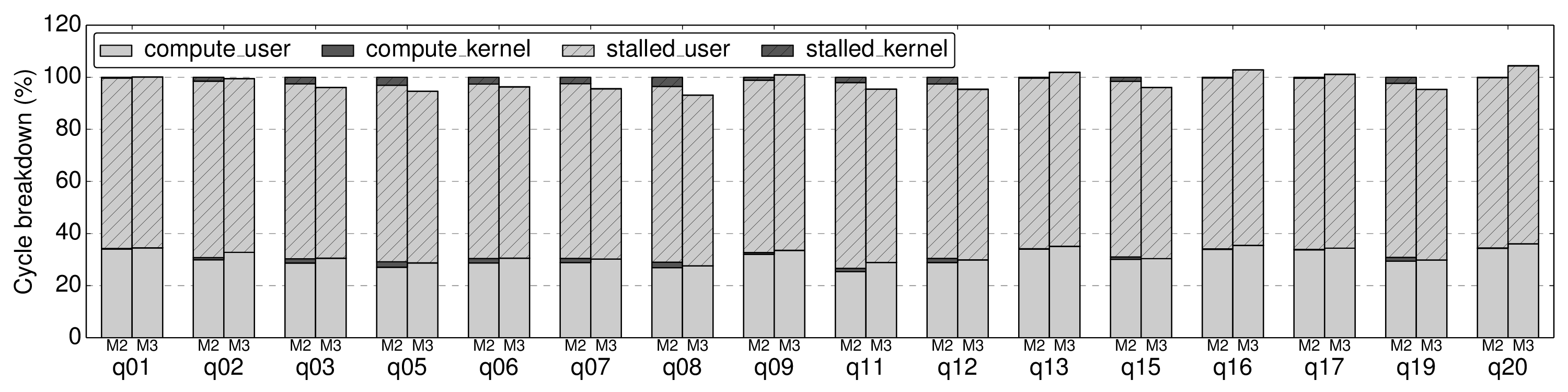}
\caption{Execution time breakdown into compute and stall cycles for 
helper thread running on same physical core as PostgreSQL in M2 and M3. Execution time is normalized with respect to M2}
\label{compute-stall-helper}
\end{figure*}

To support this explanation, Fig.~\ref{compute-stall-helper} shows the compute-stall cycle breakdown for the helper thread that shares the physical core with the computation PostgreSQL thread for \verb+M2+ and \verb+M3+ schemes. We can observe that for the queries in which \verb+M3+ performs better (e.g., Q03, Q05, Q06, and Q12), the helper thread sharing the core is more lightweight. In particular, the helper thread in \verb+M3+ does not suffer from kernel noise since the kernel related events like page faults are sorted by the other helper thread running on a different core, leading to a load reduction in the computation core. Also, a reduction in user stalled cycles can be observed (e.g., Q11 and Q15) due to lower latency memory accesses.

We find that our library is able to help improve the performance with little programming effort. The ad-hoc prefetching scheme used in \emph{pmfs\_se2}, while simple, still required code analysis to place the prefetch instructions in different places in order to maximize the number of data blocks prefetched without stalling too much the computations. With our library the creation and mapping of threads is done only once, and the creation of jobs to be enqueued came as a natural fit in a single point of the source code, i.e., when the needed block is memory mapped.

\section{Related Work}
\label{sec:RelatedWork}
Previous work on usage of NVM in context of DBMS
can be divided into three broad categories: proposals for (i)
NVM-aware DBMS designs from scratch, (ii) modification of
one or more components of an already existing in-memory
DBMS, and (iii) using NVM in a disk-oriented DBMS.

In the first category, Arulraj \textit{et al.}~\cite{arulraj2015let} propose usage of a single tier memory hierarchy, i.e., without DRAM, and compare three different storage management architectures using an NVM-only system for a custom designed lightweight DBMS. In \cite{arulraj2017build}, Arulraj \textit{et al.} discuss designing a DBMS for NVM and suggest that an NVM-enabled DBMS needs to adapt the logging protocol as well as the in-memory buffer cache in order to achieve significant performance improvements. Peloton \cite{PeletonLink} is another example of a DBMS designed from scratch for DRAM/NVM storage.

In the second category, Pelley \textit{et al.}~\cite{pelley2013storage} explore a two-level hierarchy with DRAM and NVM and study different recovery methods, using Shore-MT \cite{johnson2009shore} storage engine. Others have suggested employing NVM only for logging components of a DBMS and not for dataset storage. For example, earlier works implement NVM-Logging in Shore-MT and IBM-SolidDB~\cite{fang2011high,huang2014nvram} to reduce the impact of disk I/O on transaction throughput and response time by directly writing log records into an NVM component instead of flushing them to disk. Wang \textit{et al.}~\cite{wang2014scalable} demonstrate, by modifying Shore-MT, the use of NVM for distributed logging on multi-core and multi-socket hardware to reduce contention of centralized logging with increasing system load.

While in-memory DBMS have become quite popular, disk-based DBMS still have not lost their importance as indicated in  top ten ranking of DBMS by popularity \cite{RankingLink}. Therefore, people have investigated usage of NVM in context of disk-based DBMS. In this third category, Gao \textit{et al.}~\cite{gao2011pcmlogging} use phase changing memory (PCM) is used to hold buffered updates and transaction logs in order to improve transaction processing performance of a traditional disk-based DBMS (i.e. PostgreSQL 8.4). NVM has also been used to replace a disk-located double write buffer (DWB) in MySQL/InnoDB by a log-structured buffer implemented in NVM, resulting in higher transaction throughput \cite{son2017log}.

The work presented in this paper belongs to the third category. It focuses on usage of NVM as a replacement of disk storage in a traditional DBMS and explains necessary changes in the storage engine for such a replacement. Our contribution is an NVM-aware storage engine that is complementary to and can be applied along with PCM-logging \cite{gao2011pcmlogging} and NVM-buffering \cite{son2017log} on a traditional disk-based DBMS.



Helper threads have been used for parallel speedup of legacy applications.  Researchers have used programmer-constructed helper threads \cite{kamruzzaman2011inter} as well as compiler algorithms \cite{kim2002design} for automated extraction of helper threads to enhance the performance of applications. Use of helper-threads for loosely coupled multiprocessor systems is demonstrated in \cite{jung2006helper}, with focus on efficient thread synchronization for low overhead. Others have used special hardware to construct, spawn, optimize and manage helper threads for dynamic speculative precomputation \cite{collins2001dynamic}. Although helper-thread based prefetching is a well-studied technique, we pioneer its use in the context of NVM storage for DBMS in order to resolve the data readiness problem arising from having direct access to NVM-resident data.

\section{Conclusion}
\label{sec:conclusion}
\noindent In this paper, we study the implications of employing NVM in the design of DBMSs. We discuss the possible options to incorporate NVM into the memory hierarchy of a DBMS computing system and conclude that, given the characteristics of NVM, a platform with a layer of DRAM where the disk is completely or partially replaced using NVM is a compelling scenario. Such an approach retains the programmability of current systems and allows direct access to the dataset stored in NVM. With this system configuration in mind we modify the PostgreSQL storage engine in two incremental steps - SE1 and SE2 - to better exploit the features offered by PMFS using memory mapped I/O.

Our evaluation shows that storing the dataset in NVM instead of disk for an unmodified version of PostgreSQL improves query execution time 
by up to 39\%, with an average of 15\%. Modifications to take advantage of NVM hardware improve the execution time by around 
19\% on average as compared to disk storage. However, the current design of database software proves to be a hurdle in maximizing the improvement. 
When comparing our baseline and \textit{SE2} using PMFS, we achieve significant speedups of up to 13\% in read-dominant queries, 
but moderate average improvements of 4\% in query execution time.

We find that the limiting factor in achieving higher performance improvements is the fact that the data is not close to the processing 
units when it is needed for processing. This is a negative side effect of directly accessing data from NVM, rather than copying it into 
application buffers to make it accessible. This leads to long latency user-level cache misses eating up the improvement achieved by avoiding 
expensive data movement operations. 

We develop a general purpose data prefetching library to mitigate this negative side effect. 
The library provides services to bring data into caches through user-level helper threads in a parallel fashion without stalling the application. 
Our library improves query execution time when compared to \emph{disk\_base95} by up to 54\%, with an average of 23\%. When compared to \emph{pmfs\_base95}, query execution time improves by up to 17\%, with 8\% average improvements.

\end{document}